\documentclass{aa}  
\usepackage{natbib}
\usepackage{graphicx}
\usepackage[colorlinks=true,allcolors=blue]{hyperref}
\usepackage{xspace}
\newcommand{\SPT}{SPT0311--58\xspace}

\newcommand{\kms }{km s$^{-1}$}
\newcommand{ \Mstar } {$M_{\star}$}
\newcommand{ \Msun } {M$_{\odot}$}

\newcommand{ \Ha} {H$\alpha$}
\newcommand{ \Hb} {H$\beta$}

\newcommand{ \CII} {[C\,\textsc{ii}]}

\newcommand{\ha}{H$\alpha$}
\newcommand{\hb}{H$\beta$}

\newcommand{\sfr}{M${_\odot}$ yr$^{-1}$}

\newcommand{\OIIlines}{$[\mathrm{O}\textsc{ii}]\,\lambda\lambda 3726,29$\xspace}

\usepackage{txfonts}
\begin{document}

\title{GA-NIFS: The core of an extremely massive protocluster at the epoch of reionisation probed with JWST/NIRSpec}

\titlerunning{The core of a protocluster at z=6.9}

   \author{Santiago Arribas\inst{\ref{iCAB}}\thanks{e-mail: arribas@cab.inta-csic.es}
          \and
          Michele Perna\inst{\ref{iCAB}}
          \and
          Bruno Rodr\'iguez Del Pino\inst{\ref{iCAB}}
          \and
          Isabella Lamperti\inst{\ref{iCAB}}
          \and
          Francesco D'Eugenio\inst{\ref{iKav},\ref{iCav}}
          \and    
          Pablo G.~P\'erez-Gonz\'alez\inst{\ref{iCAB}}
          \and
          Gareth C. Jones\inst{\ref{iOxf}}
          \and
          Alejandro Crespo G\'omez
          \inst{\ref{iCAB}}
          \and
          Mirko Curti\inst{\ref{iESOge}}
          \and
          Seunghwan Lim\inst{\ref{iKav},\ref{iCav}}
          \and 
          Javier \'Alvarez-M\'arquez\inst{\ref{iCAB}}
          \and
          Andrew J. Bunker\inst{\ref{iOxf}}
          \and 
          Stefano Carniani\inst{\ref{iNorm}}
          \and 
          Stéphane Charlot\inst{\ref{iSor}}
          \and 
          Peter Jakobsen\inst{\ref{iDawn},\ref{iUCop}}
          \and
          Roberto Maiolino\inst{\ref{iKav},\ref{iCav}, \ref{iUCL}}     
          \and
          Hannah \"Ubler\inst{\ref{iKav},\ref{iCav}}
          \and 
          Chris J. Willott\inst{\ref{iNRC}}
          \and 
          Torsten B{\"o}ker\inst{\ref{iESOba}}
          \and
          Jacopo Chevallard\inst{\ref{iOxf}}
          \and 
          Chiara Circosta\inst{\ref{iESAC},\ref{iUCL}}
          \and 
          Giovanni Cresci\inst{\ref{iOAA}}
          \and
          Nimisha Kumari\inst{\ref{iSTScI}}
          \and 
          Eleonora Parlanti\inst{\ref{iNorm}}
          \and 
          Jan Scholtz\inst{\ref{iKav},\ref{iCav}}
          \and 
          Giacomo Venturi\inst{\ref{iNorm}}
          \and 
          Joris Witstok\inst{\ref{iKav},\ref{iCav}}
          }

   \institute{Centro de Astrobiolog\'ia, (CAB, CSIC--INTA), Departamento de Astrof\'\i sica, Cra. de Ajalvir Km.~4, 28850 -- Torrej\'on de Ardoz, Madrid, Spain\label{iCAB}
        \and 
        Kavli Institute for Cosmology, University of Cambridge, Madingley Road, Cambridge, CB3 0HA, UK\label{iKav}
        \and
        Cavendish Laboratory - Astrophysics Group, University of Cambridge, 19 JJ Thomson Avenue, Cambridge, CB3 0HE, UK\label{iCav}
        \and
        Department of Physics, University of Oxford, Denys Wilkinson Building, Keble Road, Oxford OX1 3RH, UK\label{iOxf}   
        \and
        European Southern Observatory, Karl-Schwarzschild-Straße 2, 85748, Garching, Germany\label{iESOge}
        \and
        Scuola Normale Superiore, Piazza dei Cavalieri 7, I-56126 Pisa, Italy\label{iNorm}
        \and
        Sorbonne Universit\'e, CNRS, UMR 7095, Institut d’Astrophysique de Paris, 98 bis bd Arago, 75014 Paris, France\label{iSor} 
        \and
        Cosmic Dawn Center (DAWN), Copenhagen, Denmark \label{iDawn}
        \and
        Niels Bohr Institute, University of Copenhagen, Jagtvej 128, DK-2200, Copenhagen, Denmark \label{iUCop}
        \and
        Department of Physics and Astronomy, University College London, Gower Street, London WC1E 6BT, UK\label{iUCL}
        \and
        National Research Council of Canada, Herzberg Astronomy \& Astrophysics Research Centre, 5071 West Saanich Road, Victoria, BC V9E 2E7, Canada\label{iNRC}
        \and
        European Space Agency, c/o STScI, 3700 San Martin Drive, Baltimore, MD 21218, USA\label{iESOba}
        \and
        European Space Agency (ESA), European Space Astronomy Centre (ESAC), Camino Bajo del Castillo s/n, 28692 Villanueva de la Cañada, Madrid, Spain\label{iESAC}
        \and
        INAF - Osservatorio Astrofisco di Arcetri, largo E. Fermi 5, 50127 Firenze, Italy\label{iOAA}
        \and 
        AURA for European Space Agency, Space Telescope Science Institute, 3700 San Martin Drive. Baltimore, MD, 21210, USA\label{iSTScI}}

\date{Received TBD; accepted TBD}

\abstract
{ The SPT0311-58 system resides in a massive dark-matter halo at z $\sim$ 6.9. It hosts two dusty galaxies (E and W) with a combined star formation rate (SFR) of $\sim$ 3500 \sfr, mostly obscured and identified by the rest-frame IR emission.
The surrounding field exhibits an overdensity of submillimetre sources, making it a candidate protocluster. }
{Our main goal is to characterise the environment and the properties of the interstellar medium (ISM) within this unique system.}
{We used spatially resolved low-resolution ($R$=100) and high-resolution ($R$=2700) spectroscopy provided by the JWST/NIRSpec Integral Field Unit to probe a field of $\sim$ 17 $\times$ 17 kpc$^2$ around this object, with a spatial resolution of $\sim$ 0.5 kpc.} 
{These observations reveal ten new galaxies at z$\sim$ 6.9 characterised by dynamical masses spanning from $\sim$10$^{9}$ to 10$^{10}$ \Msun\ and a range in radial velocity of $\sim$ 1500 \kms, in addition to the already known E and W galaxies. The implied large number density ($\phi$ $\sim$ 10$^{4}\ $Mpc$^{-3}$) and the wide spread in velocities confirm that \SPT is at the core of a protocluster immersed in a very massive dark-matter halo of  $\sim$ (5 $\pm$ 3) $\times$ 10$^{12}$ \Msun, and therefore represents the most massive protocluster ever found at the epoch of reionisation (EoR). We also studied the dynamical stage of 
its core and find that it is likely not fully virialised.  
The galaxies in the system exhibit a wide range of properties and evolutionary stages. The contribution of the ongoing H$\alpha$-based unobscured SFR to the total star formation (SF) varies significantly across the galaxies in the system.
Their ionisation conditions range from those typical of the field galaxies at similar redshift recently studied with JWST to those found in more evolved objects at lower redshift, with log([OIII]/\Hb) varying from $\sim$ 0.25 to 1. The metallicity spans more than 0.8 dex across the FoV, reaching nearly solar values in some cases. The detailed spatially resolved spectroscopy of the E galaxy reveals that it is actively assembling its stellar mass, showing inhomogeneities in the ISM properties at subkiloparsec scales, and a metallicity gradient ($\sim$ 0.1 dex/kpc) that can be explained by accretion of low metallicity gas from the intergalactic medium. The kinematic maps also depict an unsettled disc characterised by deviations from regular rotation, elevated turbulence, and indications of a possible precollision minor merger.}
{These JWST/NIRSpec IFS observations confirm that \SPT is at the core of an extraordinary protocluster, and reveal details of its dynamical properties. They also unveil and provide insights into the diverse properties and evolutionary stages of the galaxies residing in this unique environment.}


\maketitle 
\section{Introduction}
\label{sec: Intro}

In the local Universe, observations have established the presence of massive virialised clusters of galaxies formed by hundreds to thousands of mainly passively evolving red quenched galaxies (\citealp{Dressler1980}). Cosmological models based on the $\Lambda$CDM paradigm explain these local clusters as the hierarchical build-up of overdensities in the early Universe (\citealp{Springel_2005}; \citealp{Overzier2009}). Protoclusters represent the early nonvirialised stages of the galaxy cluster assembly, a phase when the galaxies were actively building up their stellar component and the associated processes were 
in full action (i.e. gas inflow from the cosmic web, supermassive black hole coevolution,  metal enrichment of the intracluster media, mergers and interactions, etc.). The protocluster phase therefore plays a crucial role in both testing cosmology and studying galaxy formation and evolution  (\citealp{Overzier2016}; \citealp{Alberts2022}).

Considerable effort has been dedicated in recent decades to the search for clusters and protoclusters at increasing redshifts. At z > 2-3, the search techniques typically employed at lower redshifts become inefficient, as the cluster members are difficult to distinguish photometrically from those in the field, and the hot intracluster medium (ICM) gas signatures (i.e. X-ray emission and Sunyaev-Zel'dovich effect; e.g. \citealp{Mantz2014}; \citealp{Bleem2015}) are out of reach with current instrumentation. The presence of protoclusters at early cosmic times is therefore generally identified by a large concentration of galaxies in angular coordinates and redshift (c.f. \citealp{Overzier2016}). 

An efficient way to search for overdensities at high-z is to probe the environment of objects that are known to be the progenitors of the very massive brightest cluster galaxies (BCGs) that reside at the centre of local clusters. One such class is the high-z dusty star-forming  galaxies (DSFGs; 
\citealp{Casey2014,Casey2021};  \citealp{Swinbank2014};  \citealt{Magnelli2013}), which are thought to be the progenitors of local massive ellipticals (\citealp[e.g.][]{Simpson2014}) and quiescent galaxies at intermediate redshifts (z $\sim$ 2-4; e.g. \citealp[]{Toft2014}). Hence, several studies have highlighted their importance for identifying overabundances at high-z and studying the protocluster phase 
(e.g. \citealp{Chapman2001,Chapman2009}; \citealp{Daddi2009}; \citealp{Capak2011}; \citealp{Walter2012}; \citealp{Casey2014}; \citealp{Dannerbauer2014}; \citealp{Casey2016}; \citealp{Wang2016}; \citealp{Harikane2019}; \citealp{Wang2021}; \citealp{Lim2021}; \citealp{Calvi2023}; \citealp{Zhou2023}).

The advent of the {\it James Webb} Space Telescope (JWST; \citealp{Gardner2023}), with its orders of magnitude improvement in sensitivity, enhanced angular resolution in near- and mid-infrared wavelengths, and superb instrumentation, is now opening the opportunity to study both the protocluster phase and the nature of high-z DSFGs in unprecedented detail. Recent studies using a variety of JWST observing modes have revealed galaxy overdensities around the DSFG HDF850.1 at z = 5.2 (\citealp{Herard-Demanche2023}; \citealp{Sun2024}), the QSOs J0100+2802 at z = 6.3 (\citealp{Kashino2023}), and J0305–3150 at z = 6.6 (\citealp{Wang2024}) behind cluster SMACS0723$-$7327 at z=7.66 (\citealp{Laporte2022}) and A2744-z7p9OD at z = 7.88 (\citealp{Morishita2023}; \citealp{Hashimoto2023}). 
Moreover, protoclusters have also been identified in several regions within the GOODS-North and GOODS-South fields with redshifts ranging from 5.2 to 8.2 (\citealp{Helton2023a,Helton2023b}), and a protocluster core candidate has been found around GN-z11 at z=10.6 (\citealp{Scholtz2023_astroph}).  Recent JWST studies have also unveiled the properties of the DSFGs and their contribution to the star formation activity at cosmic dawn (e.g. \citealp{perez-gonzalezCEERSKeyPaper2023}, \citealp{Zavala2023}, \citealp{Akins2023}, \citealp{Barrufet2023}). 

In this work, we present JWST Near-Infrared Spectrograph (NIRSpec) Integral Field Unit (IFU) observations of SPT-S J031132-5823.4 (hereafter \SPT) at z=6.90, the most distant and massive DSFG known (\citealp{Strandet2017}). ALMA observations by \cite{Marrone2018} revealed that this mildly lensed system ($\mu$ $\sim$ 1.3-2.2) consists of a pair of massive dusty galaxies separated by a projected distance of 8 kpc, that form stars at extreme rates (SFR $\sim$ 540 and 2900 \sfr) and contain enormous amounts of dust (M$_{dust}$ $\sim$ 1-3 $\times$ 10$^{9}$ \Msun) and gas (M$_{gas}$ $\sim$ 1-3 $\times$ 10$^{11}$ \Msun; see also \citealp{jarugula_molecular_2021}, \citealp{Witstok2023}). These galaxies have a clumpy structure and turbulent kinematics (\citealp{Spilker2022}; \citealp{Alvarez-MarquezSPT}), and are undergoing fast metal enrichment (\citealp{Litke2023}). \cite{Marrone2018} also show that \SPT resides in one of the most massive dark-matter halos expected to exist at z$\sim$ 7.  Due to its extreme luminosity at submillimetre (submm) 
wavelengths, \SPT was also included in the study by \cite{Wang2021}, who searched for overdensities in the fields around the brightest high-redshift sources discovered in the SPT 2500 deg$^2$ survey (\citealp{Vieira2010}, \citealp{Mocanu2013}, \citealp{Everett2020}). This study found that the surroundings of \SPT (i.e. R $\lesssim$ 1.3 Mpc)  show a significant overdensity of submm sources and a compacted central distribution, suggesting that it could potentially be at the centre of one of the most distant protoclusters identified so far. In summary, all these previous works have established that SPT0311-58 is indeed an extraordinary system that can be seen as an ideal laboratory for studying galaxy evolution in a massive halo at the epoch of reionisation (EoR). 

The JWST/NIRSpec IFU observations presented here cover the rest-frame UV wavelengths and main optical emission lines of \SPT (i.e. $\sim$ 0.13 -- 0.67 $\mu$m). This allows us to characterise the physical and kinematic properties of the warm ionised gas component of its ISM at subkpc scales (i.e. $\sim$ 0.5 kpc) over a field of view of $\sim 17 \times 17 $\,kpc$^2$. Being at z=6.9, \SPT is at about the highest redshift for which H$\alpha$ can be observed with the high angular resolution and sensitivity provided by JWST/NIRSpec. 

The present paper is organised as follows. In Section \ref{sec: Obs_and_reductions}, we describe the observations and reductions.  In Section \ref{subsec:Characterization}, we explain how we identified and spectrally characterised a group of ten newly discovered z $\sim$ 6.9 galaxies around the two already known massive galaxies in \SPT,  together with other lower z sources also found in the field of view (FoV).  
Section \ref{sec:ISMconditions} presents the stellar masses and then focuses on the ISM conditions, including the ionisation and metallicity properties. We also obtain and discuss the ongoing H$\alpha$-based (unobscured) star formation. In Sect. \ref{subsec:kinematicsanddynamics}, we discuss the kinematic properties of the system. In light of the results of this study, in Sect. \ref{Sec:Discussion}, we discuss the protocluster scenario for \SPT.  We also discuss the physical, chemical, 
and kinematic properties of the E galaxy, for which the IFU data give detailed 2D information. Finally, in Sect. \ref{section:summary}, we summarise the main results of this study. Appendix \ref{appendix:spectral_methods} provides details of the spectral analysis and Appendix \ref{appendix:lens_model} presents the magnification map based on a gravitational lens model for the SPT0311-58 system, making use of the new information provided by these IFU data.  

In this work, we adopt the cosmological parameters from \cite{Planck:2015}: $H_0$ =  67.7 \kms Mpc$^{-1}$, $\Omega_m$ = 0.307, and $\Omega_\Lambda$ = 0.691, which provide a scale of 5.41 $ \mathrm{kpc}/\arcsec$ at $z=$6.90. Unless otherwise stated, we use proper distances throughout the paper. We refer to the two main galaxies in SPT0311-58 as W and E, according to their relative position on the sky. We adopt a Kroupa initial mass function (IMF; \citealp{Kroupa2001}) 

\section{Observations and reductions}
\label{sec: Obs_and_reductions}

\subsection{SPT0311-58 observations}
\label{subsec: observations}

The present work is part of the GA-NIFS (Galaxy Assembly with NIRSpec Integral Field Spectroscopy), a NIRSpec GTO program designed to study the internal structure and the environment of a sample of $\sim$ 55 galaxies and active galactic nuclei (AGN) with redshifts of between 2 and 11 by means of spatially resolved spectroscopy obtained with IFS mode (\citealp{BoekerIFS}) of the NIRSpec instrument (\citealp{Jakobsen2022}). Details of the GA-NIFS program can be found in \cite{Perna2023IAU}.  Among other subsamples, the program includes several distant (i.e. z $>$ 4) infrared-selected dusty star-forming galaxies: HLFS3 (\citealp{Jones23}),  ALESS073.1 (\citealp{Parlanti2023}), GN20 (\citealp{Ubler2024}), and SPT0311-58, subject of the present study. 

The NIRSpec-IFU observations of SPT0311-58 were executed in November 20, 2022, and are specified in PID1264 (PI: L. Colina). These NIRSpec observations were combined in a single proposal with independent MIRI observations of the same target (presented in \citealp{Alvarez-MarquezSPT}) with the aim of saving the telescope slew overhead. The NIRSpec observations include two different spectral configurations: low-resolution (R100) spectroscopy with the prism covering the full 0.6--5.3 $\mu$m range, and high-resolution (R2700) spectroscopy with the grating/filter combination G395H/F290LP, which covers from 2.8 to 5.27 $\mu$m, though due to the gap between the two detectors the $\sim$ 3.98–4.20 $\mu$m range is not observed at R2700 (\citealp{BoekerIFS}). The FoV is $\sim$ 3$''$ $\times$ 3$''$ (i.e. $\sim$ 17 $\times$ 17 kpc$^2$ at the redshift of \SPT), and the native spaxel size 0.1$''$ (0.54 kpc).  

We constrained the allowed position angle range for the observations to minimise the presence of bright sources in the MSA FoV that could leak light and contaminate the IFU spectra.  We were primarily concerned about this potential problem for the R100 spectra, and therefore we protected the region of the detector occupied by these spectra. For the high-resolution spectra, MSA-leakage is not so critical as our main goal with this mode is to analyse emission lines. 

We use the first four positions of the medium cycling dither pattern, which provide offsets large enough to deal with the failed open micro-shutters and other sources of background, while keeping a large effective FoV with the complete exposure time. Also, this choice minimises the number of individual exposures, and therefore also the (dominant) detector noise.  

No dedicated background exposures were obtained, as it was expected that a substantial number of spaxels were free from galaxy emission and suitable for deriving the background spectrum. 

The detector was configured with the improved reference sampling and subtraction pattern (IRS$^2$), which significantly reduces the readout noise with respect to the conventional method \citep{RauscherIRS2}. In particular,  NRSIRS2RAPID (60 groups) and NRSIRS2 (25 groups) options were respectively selected for R100 and R2700 in a single integration. The different temporal sampling between the two spectral configurations was motivated by data volume constraints. This gives a total on-source integration time of 3559 and 7352 s for R100 and R2700, respectively.  
  
\subsection{NIRSpec IFU data reduction} 
\label{subsec:reductions}
The data reduction was performed with the JWST calibration pipeline (\citealp{bushouse_howard_2023_8380331}) version 1.8.2, using the context file {\it jwst\_1068.pmap}. We follow the general procedure to correct detector level defects (Stage 1), calibrate the data (Stage 2), and combine the individual exposures and build the datacube (Stage 3). However, several modifications to the default reduction steps were required to enhance data quality. These modifications, described in detail in \citet{Perna_LBQS}, are briefly summarised here.

\begin{itemize}

\item The individual count-rate frames were further processed at the end of Stage 1 to correct for the different zero-level in the individual (dither) frames and subtract the vertical stripes induced by the 1/f noise. 

\item We masked the regions affected by leakage from failed open microshutters, including not only the pixels flagged as ‘MSA$\_ $FAILED$\_ $OPEN', but also others not correctly associated with such leakage in the context file. 

\item In Stage 2, we masked detector pixels corresponding to the edges of the IFU slices with unreliable {\it sflat} corrections, as they cause evident artefacts (stripes) in the final data cube.    

\item Outliers were flagged and removed on the individual 2D exposures at the end of Stage 2, using an algorithm similar to {\sc lacosmic} \citep{vandokkum2001}. We calculated the derivative of the count-rate maps along the (approximate) dispersion direction, and rejected all measurements above a given percentile of the resulting distribution (\citealp{D'Eugenio_rotator_PSF}).  This threshold value was optimised after inspecting the products of the different realisations. We found that 98{th} and 99.5{th} percentiles were appropriate in our case for the R100 and R2700 data, respectively.

\item {After experimenting with the different methods and scale options, we generated the final datacubes with the drizzle weighting method and a spaxel scale of 0.05\arcsec in the \it cube\_build} step.

\end{itemize}

As it has been reported by several authors (e.g. \citealp{RodriguezdelPino2023arXiv}), the noise in the datacube provided by the pipeline (i.e. `ERR' extension) is underestimated, and we therefore rescaled it according to the standard deviation of the data themselves (see Sect. \ref{subsec:Stellar masses}).

\begin{figure*}
    \centering
    \includegraphics[width=0.85\textwidth]{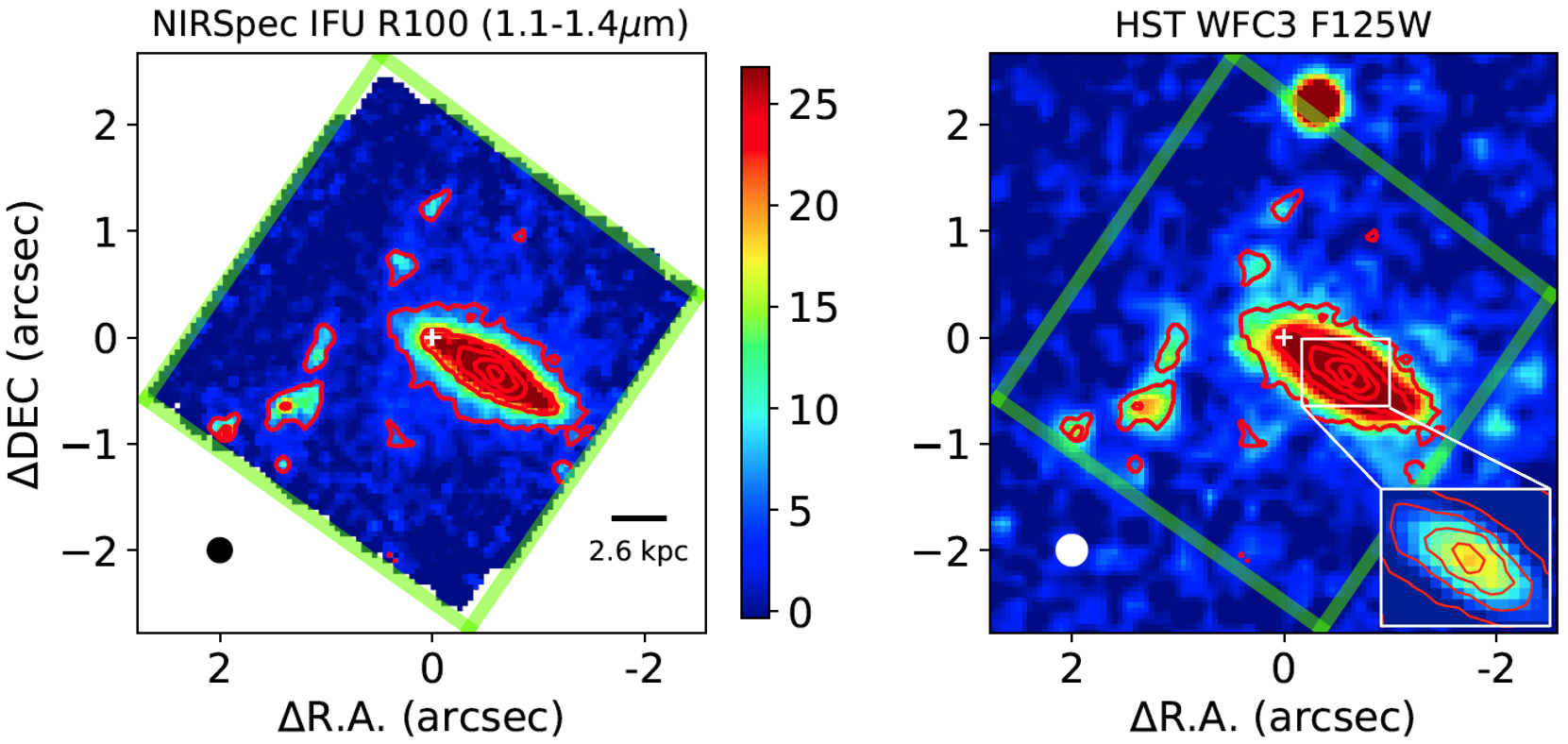}
    \caption{Comparison of JWST/NIRSpec IFU and HST imaging. 
    Left: Image generated from the NIRSpec R100 datacube, integrating over the spectral range 1.1-1.4 $\mu$m. The colour bar is in units of 10$^{-19}$ erg s$^{-1}$cm$^{-2}$ per spatial pixel. Right: HST image obtained with WFC3 and F125W filter resampled at the pixel size used with NIRSpec (i.e. 0.050 \arcsec). The inset zooms in on the region around the emission peak used to align the NIRSpec cube (red isophotes) with the image represented with an unsaturated colour code. Here, and in all the spectral maps presented in this work, the (0,0) corresponds to coordinates $\alpha$ = 03$^h$ 11$^m$ 33$^s$.248 and $\delta$ = --58º 23' 33''.24 (white cross). In the image the peak of the lens galaxy is at relative coordinates (--0.64 \arcsec, --0.29 \arcsec). The green lines indicate the border of the NIRSpec IFU FoV.  The black and white circles in the bottom-left corners of the panels represent the FWHW of the respective PSFs. 
    } 
    \label{fig:F125W}
\end{figure*}


\subsection {Astrometric registration}
\label{subsec:Astrometric}

In order to set the reduced data cube coordinates onto the \textit{Gaia} reference system, we applied the following two-step procedure.  Firstly, we aligned the IFU data with the \textit{Hubble} Space Telescope (HST) WFC3 images. With this goal, we generate an image over a band-pass similar to F125W by collapsing the R100 data-cube in the spectral range 1.1-1.4 $\mu$m (see Fig. \ref{fig:F125W}). Then we obtained the coordinate offsets to be applied to align this image with the F125W HST image, taking as reference the flux peak of the bright foreground galaxy (see inset in Fig. \ref{fig:F125W}). We estimate the uncertainties associated with this step to be $\lesssim$ 70 mas on each axis.  In this figure we can appreciate the slightly better resolution provided by NIRSpec IFU with respect to HST at these wavelengths, as expected from their respective PSFs (NIRSpec: \citealp{D'Eugenio_rotator_PSF}; WFC3: \citealp{Dressel2023}). 
Secondly, the absolute astrometry of the HST/WFC3 F125W image was derived by matching the position of four stars present in the image FoV with the \textit{Gaia} DR3 catalogue \citep{GAIA2022}. The absolute sky coordinates (ICRS) of the centre of the FoV (i.e. spaxel [55,55]) are $\alpha$ = 03$^h$ 11$^m$ 33$^s$.248 and $\delta$ = --58º 23' 33''.24, and correspond to (0,0) in Fig. \ref{fig:F125W}, and other spectral maps in the paper. 
The uncertainty in the astrometric alignment is 17~mas in R.A. and 30~mas in Dec. The total offsets applied were 0.155 and 0.305 arcsec in R.A. and Dec., respectively, and the absolute coordinates uncertainty, taking into account both steps, are estimated to be $\sim$ 80 mas in each direction. 

We checked that the images generated from the R100 and R2700 data-cubes were consistently aligned with each other. In fact, a 2D Gaussian fit to the brightest region of the lens galaxy in a common wavelength range led to differences smaller than a fraction of a spaxel (i.e. $<$ 10 mas).    

\subsection {Background}
\label{subsec:Background}
\begin{figure*}[ht]
    \centering
    \includegraphics[width=15.4cm,trim= 0 0 0 0,clip]
    {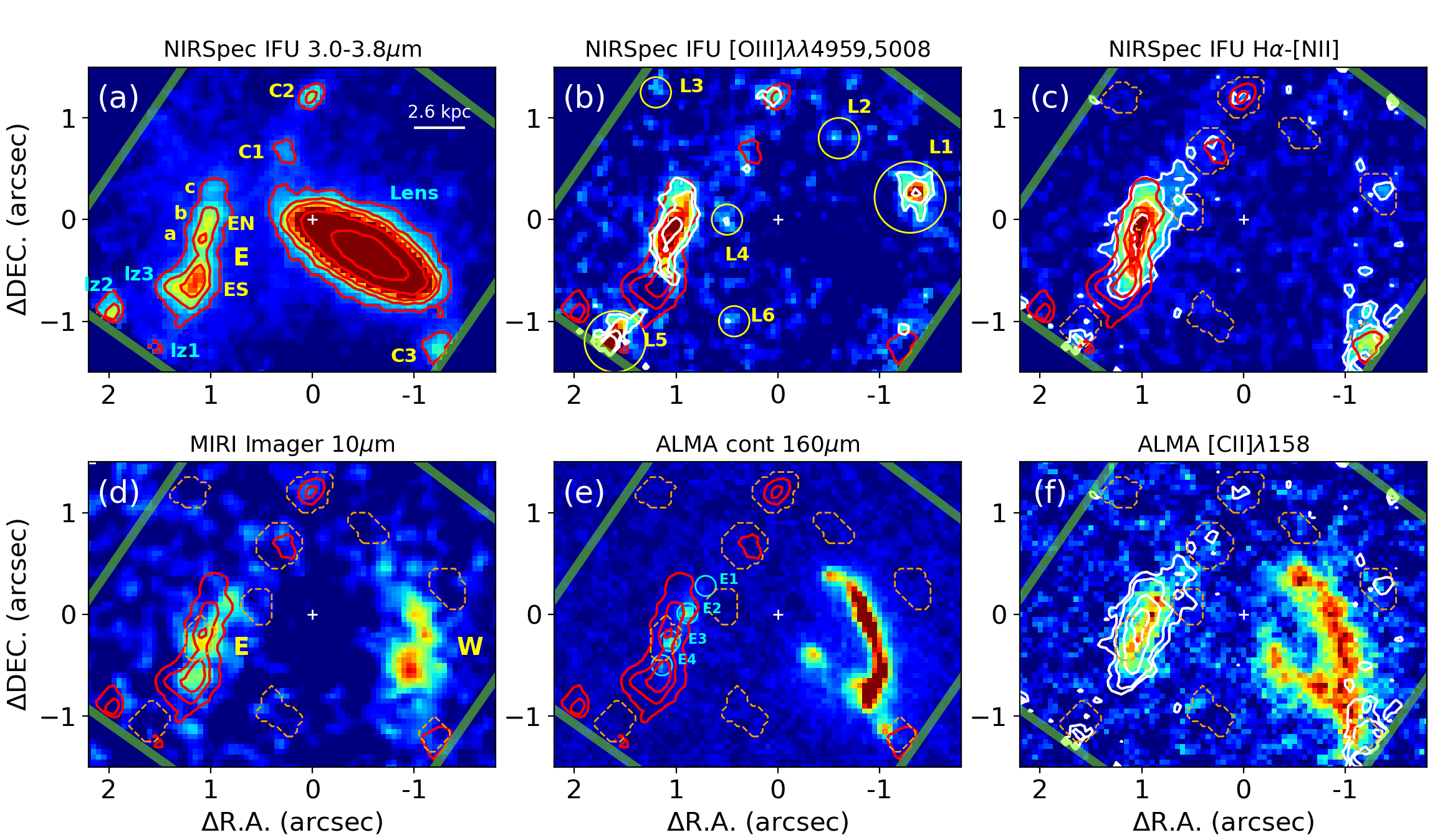}
    \caption{Continuum and line emission NIRSpec IFU images, together with MIRI and ALMA data. (a): 
    NIRSpec image generated by collapsing the R100 datacube in the 3.0-3.8 $\mu$m spectral range (see text). The continuum emitting sources, whose spectra indicate they are at $z$ $\sim$ 6.9, are identified with yellow labels  (i.e. C1, C2, C3, E). In the E galaxy, we distinguish several subregions (ES, ENa,b,c). Sources that according to their spectra are at lower z (i.e. lz1, lz2, lz3) are labelled in cyan, along with the foreground lens galaxy.  The red isocontours (in red) correspond to 5, 10, 15, 30, and 100 $\sigma$ values.  
    (b): [OIII] emission image at z $\sim$ 6.9 obtained from the R100 datacube (see text for specific spectral ranges).  Six line-emitting sources that were not detected in the continuum map (panel a), are identified in this map as z $\sim$ 6.9 sources and labelled in yellow (L1 to L6). White isolines correspond to 3, 5, 10, and 15 $\sigma$. Here, and in panels (c), (d), and (e), the red contours are the same as in panel (a), after removing the ones for the lens galaxy for clarity. 
    (c): Same as panel (b) but for the H$\alpha$ emission.
    (d): MIRI image at 10 $\mu$m, after subtracting the lens (\citealp{Alvarez-MarquezSPT}). 
    (e): ALMA continuum at 160 $\mu$m. We mark clumps E1 -- E4 identified by \cite{Spilker2022}. 
    (f): ALMA map of the [CII]$\lambda$158 line. The white isolines correspond to the H$\alpha$ emission. 
    Fig. \ref{fig:schematic_apertures.png} identify the targets and their apertures (for details see Appendix \ref{appendix:spectral_methods}). In panels (c-f), these apertures are also shown with orange dashed lines.
    The green straight lines in all the pannels indicate the border of the NIRSpec IFU FoV.}

    \label{fig:R100imaging}
\end{figure*}

\begin{figure}[h]
    \centering
    \includegraphics[width=0.49\textwidth]{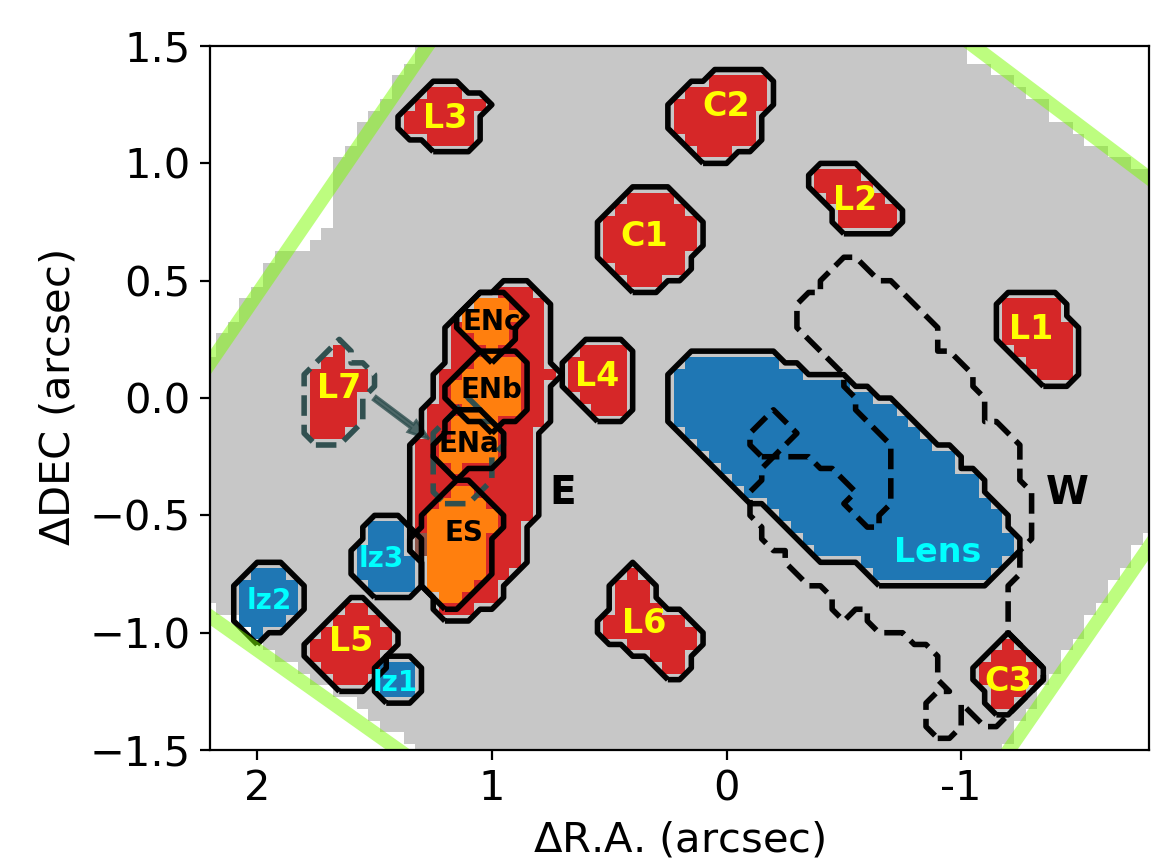}
    \caption{Identification of the different targets and regions in the FoV of \SPT. The apertures used to extract the spectra of the $\sim$ 6.9 sources are coded in red (with the subregions within E in orange), and those for the low-z galaxies in blue. Source L7, which is identified by spectral decomposition (Sect. \ref{subsubsection:Decomposition_twocomponentmaps}), is shown for completeness. The W galaxy is not detected at the resolution of the IFU, and is shown with a dashed line. More details about the apertures in Appendix \ref{appendix:spectral_methods}.}   
    \label{fig:schematic_apertures.png}
\end{figure}

For the R100 datacube, the background spectrum was obtained from the median of about 1550 spectra free from emission coming from \SPT, the lens galaxy, or any other object in the FoV. These spaxels were identified by inspecting the continuum and the line maps generated across the whole R100 wavelength range (Sect. \ref{subsec:Characterization} and Fig. \ref{fig:R100imaging}) to make sure that neither continuum nor line emission from the galaxies was included. The area covered by the background spaxels is shown in Fig. \ref{fig:Apertures}, which avoids the regions with emission from the galaxies (Fig. \ref{fig:schematic_apertures.png}).

We report here that the individual R100 spectra presented a strong feature at 1.083 $\mu$m, which may be due a metastable He line (\citealp{Brammer2014}). This feature is found over the whole FoV of the IFU with a uniform intensity. It is visible in the spaxels used to obtain the background spectrum (see Fig. \ref{fig:R100low-z} bottom panel). It is already present in the raw data, and also in the reference file {\it nirspec$\_$sflat$\_$0185}. As for the present analysis, we note that this emission is removed in the background subtraction step. 

For the R2700 spectra, we do not subtract the sky background. These high-resolution data are only used for the study of the emission lines and the mean continuum is removed during the fitting process (see Appendix \ref{appendix:spectral_methods}).

\subsection {Ancillary data}
\label{subsec:ancillary}
In addition to the NIRSpec IFU data presented above, in this study we made use of additional data that are briefly described below. 

\begin{itemize}

\item  JWST/NIRSpec IFU: Calibration star: To derive aperture corrections (Appendix \ref{subsubsec:aperture_corrections}) and the FHWM of the PSF, we used the high-resolution G395H/290LP observations of the standard star 1808347 (PID:1128, Obs. 9; PI: N. Luetzgendorf), which were reduced with the same pipeline version and context file as the data for SPT0311-58.
  
\item HST imaging: HST imaging of \SPT obtained with the WFC3 (filter F125W) was used to set the coordinate system of the present NIRSpec data into \textit{Gaia} (see Sect. \ref{subsec:Astrometric}). The image was directly downloaded from MAST (PID14740; PI: D. Marrone).  

\item ALMA: We used ALMA band 6 observations of the \CII158$\mu m$ line and the adjacent continuum from programs 2016.1.01293.S and 2017.1.01423.S  (P.I.  D.  Marrone).  
These data have been presented in \citet{Marrone2018} and \citet{Spilker2022}. We requested the calibrated measurement sets (MS) from the EU ALMA Regional Centre (ARC).  Data were calibrated with the standard pipeline procedure. To analyse the data, we used the ALMA data reduction software {\tt CASA} v6.2.1 \citep[][]{CASA}. We created a map of the continuum combining two line-free spectral windows, obtaining a synthesised beam with FWHM of $0.09\times 0.08$~arcsec$^2$ and a sensitivity of $9.4$~$\mu$Jy~beam$^{-1}$. As for the \CII\ data-cube, the synthesised beam FWHM is $0.08\times 0.07$~arcsec$^2$ with a sensitivity of $\sim 150$~$\mu$Jy~beam$^{-1}$ in a 20~\kms channel. The \CII\ moment 0 map (intensity map) was generated by integrating the continuum subtracted data-cube in the spectral channels from -1000~\kms\ to 1200~\kms\ with respect to $z=6.900$. Before producing the moment 1 and moment 2 maps, the spaxels with low signal-to-noise ratio (S/N< 2.3)  and those with spurious emission in each 20~\kms\ velocity channel were masked.  
Finally, the continuum image and \CII\ data-cube were resampled to a spaxel size of 0.05" to match the size of the NIRSpec spaxels using the {\tt CASA}  task {\tt imregrid}. 

\item JWST/MIRI: The image of SPT0311-58 obtained with MIRI at 10 $\mu$m and published by \cite{Alvarez-MarquezSPT} was kindly provided to us by the MIRI GTO team.  

\end{itemize}

\section{Field characterisation: Discovery of 10 new z\texorpdfstring{$\sim$}{=}6.9 galaxies (and 3 low-z objects)}
\label{subsec:Characterization}

The field around SPT0311-58 is very complex, as it includes a large and bright lens galaxy, the two previously identified main galaxies (i.e. E and W) and, as detailed below, ten additional newly discovered galaxies at z $\sim$ 6.9, as well as a further three lower redshift objects. In this section, we identify all of these sources in the FoV, in most of the cases through spectral imaging with the NIRSpec R100 data cube. We then present their R2700 and R100 spectra, which allow us to confirm their redshifts. We also briefly compare the NIRSpec images with those obtained with MIRI and ALMA at longer wavelengths. Finally, we  identify one galaxy at z $\sim$ 6.9 via two-dimensional spectral decomposition of the R2700 spectra.      

\subsection{NIRSpec R100 spectral imaging: Identification of continuum and line-emitting sources at z\texorpdfstring{$\sim$}{=} 6.9 }
\label{subsubsec:identification}

Fig. \ref{fig:R100imaging} (a) shows the image obtained from the R100 data-cube after integrating over the spectral range 3.0 -- 3.8 $\mu$m. For z $\sim$ 6.9 sources, this spectral range is dominated by the stellar continuum, as it does not include prominent lines (i.e. rest-frame $\sim$ 0.38 -- 0.48 $\mu$m). 
The image shows the very bright lens galaxy, the E galaxy, and several continuum-emitting galaxies labelled as lz1, lz2, lz3, C1, C2, C3. The structure of the E galaxy is clearly resolved, allowing us to distinguish the south and north regions (ES and EN, respectively), which are also clearly seen in Fig. \ref{fig:F125W}. We also define subregions within EN with lower case letters (i.e. a, b, c, from south to north). 
As we will see  below (Sect. \ref{subsubsec:lens,and other low-z}), the spectra of the sources lz1, lz2, and lz3 indicate that they are at lower z than \SPT (i.e. z < 3,  hence the `lz' identification label). Together with the lens galaxy, all these low-z objects are labelled in cyan in the figure. One of these low-z objects (lz3) partially overlaps with ES.  The spectra of sources C1, C2, and C3 (Sect. \ref{subsubsec:spectra_high-z }) indicate that they are at z$\sim$ 6.9, and together with other sources at this redshift (i.e. E and its substructures) they are labelled in yellow in Fig. \ref{fig:R100imaging} (a). To ease the identification of sources detected in the field we include the apertures used for extracting their spectra in Fig. \ref{fig:schematic_apertures.png}.  

Fig. \ref{fig:R100imaging} (b) and (c) presents the [OIII]$\lambda$$\lambda$4959,5008 and H$\alpha$ (+[NII]$\lambda$$\lambda$6548,6584) emission line maps, obtained by integrating the spectral ranges that cover these lines at a redshift of $\sim$ 6.9, and subtracting a continuum measured close to the lines\footnote{Specifically, the selected ranges are: 5.00 -- 5.15 $\mu$m for H$\alpha$ continuum and 5.17 -- 5.23 $\mu$m for the H$\alpha$ line, and 3.72 -- 3.82 $\mu$m for the [OIII] continuum and 3.90 -- 4.00 $\mu$m for the [OIII]$\lambda$$\lambda$4959,5008 lines.}. Hence, these images allow us to identify line-emitting sources at that particular redshift. These images reveal a total of nine line-emitters: the continuum-detected C1, C2, C3 and six new sources L1-L6 (see also Fig. \ref{fig:schematic_apertures.png}). The detection of several emission lines in the spectra of all these sources confirmed that they are at a redshift of $\sim$ 6.9 (see below in Sect. \ref{subsubsec:spectra_high-z } ).    
 
We note the morphological differences between the continuum and the line maps. For instance, some bright line emitting sources, like L1 and L5, are barely detected in the continuum. [OIII] and \ha\ images also show distinct morphology. The E galaxy is more extended in H$\alpha$ than in [OIII], and shows a faint diffuse component towards the north-west. The [OIII] shows a tail towards the south-west, which is not seen in \ha.  Source C3 is much brighter in H$\alpha$ than in [OIII], and to the north there is also a zone of faint extended \ha\ emission.

The top panels in Fig. \ref{fig:R100imaging} show that the main structure of the W galaxy is not detected at spaxel level in the NIRSpec images (continuum, \ha, or [OIII]). However, as shown in Sect. \ref{subsubsec:spectra_high-z }, the R100 spectrum obtained integrating over a large aperture enables us to recover its \ha\ flux.

\subsection{Morphological comparison with MIRI and ALMA imaging}
\label{subsubsec:comparisonMIRI_ALMA}

Fig.\ref{fig:R100imaging}(d) shows the 10 $\mu$m image obtained with JWST/MIRI, after subtracting the lens foreground galaxy (\citealp{Alvarez-MarquezSPT}). At z=6.9, this image traces the stellar-continuum at the near-IR rest-frame (i.e. 1.26 $\mu$m). The comparison with the NIRSpec 3.0-3.8 $\mu$m image (Fig.\ref{fig:R100imaging}(a), rest-frame $\sim$ 0.43 $\mu$m) shows clear differences. Although in the central regions of the E galaxy the bulk of the NIRSpec emission approximately matches the MIRI image, in the northern zone the mid-infrared emission seems more extended to the west. The emission detected with MIRI in the W galaxy has no NIRSpec counterpart. Conversely, none of the new sources at z $\sim$ 6.9 discovered with NIRSpec were reported by \cite{Alvarez-MarquezSPT} as MIRI emitters. In fact, we performed aperture photometry at their positions and find that none of them have 10 $\mu$m flux above the 5-sigma noise level (0.345 $\mu$Jy).   

The NIRSpec and ALMA continuum images also show very different structures. The W galaxy, very bright at 160 $\mu$m (Fig. \ref{fig:R100imaging}(e)), is not seen by NIRSpec in the 3.0-3.8 $\mu$m map (Fig. \ref{fig:R100imaging}(a)).  The NIRSpec region ENa and the ALMA clump E3 are in positional agreement within $\sim$ 0.1 arcsec. 
None of the other NIRSpec sources at z $\sim$ 6.9 have ALMA continuum detections, except C3 whose aperture partially overlaps with the southern structure of W. 

As for the line emission maps, the main body of E in H$\alpha$ and  [OIII] (Fig. \ref{fig:R100imaging}(b,c)) approximately matches the [CII]$\lambda$158 $\mu$m emission (Fig. \ref{fig:R100imaging}(f)). Nearby regions C1 and L4 also have low [CII] emission. Regarding the W galaxy, the bright [CII] regions are not seen with NIRSpec, except weakly at L1. In addition, the C3 aperture overlaps with the main W emission, but the fact that it is at the edge of the NIRSpec FoV makes the comparison difficult. The other NIRSpec newly discovered sources at z $\sim$ 6.90 are not detected with ALMA at [CII]$\lambda$158 $\mu$m.

\subsection{Spectral characterisation of SPT0311-58}
\label{subsec: Spectral_caracterization}

To study the spectral properties in the complex FoV of \SPT, we followed different approaches.  In this section, we first present the integrated spectra extracted from the data cubes for the sources and regions identified above. For details on the spectral analysis see Appendix \ref{subsec:extractionapertures}. As we describe below, these spectra show that three objects (i.e. lz1, lz2, and lz3) are at low redshift, while confirming that nine newly discovered sources are at z $\sim$ 6.9. 

Later in this section, we present a spatially resolved analysis of the E galaxy using the R2700 data. This analysis reveals an additional distinct galaxy at z$\sim$ 6.90 overlapping along the line of sight of E. In summary, we  find ten galaxies at redshift $\sim$ 6.9 (nine sources identified through R100 spectral imaging, and one through the analysis of the R2700 data), in addition to the already known E and W galaxies.

\subsubsection{Spectra of the lens and other low-z galaxies}
\label{subsubsec:lens,and other low-z}

\begin{figure}[h]
    \centering
    \includegraphics[width=0.49\textwidth]{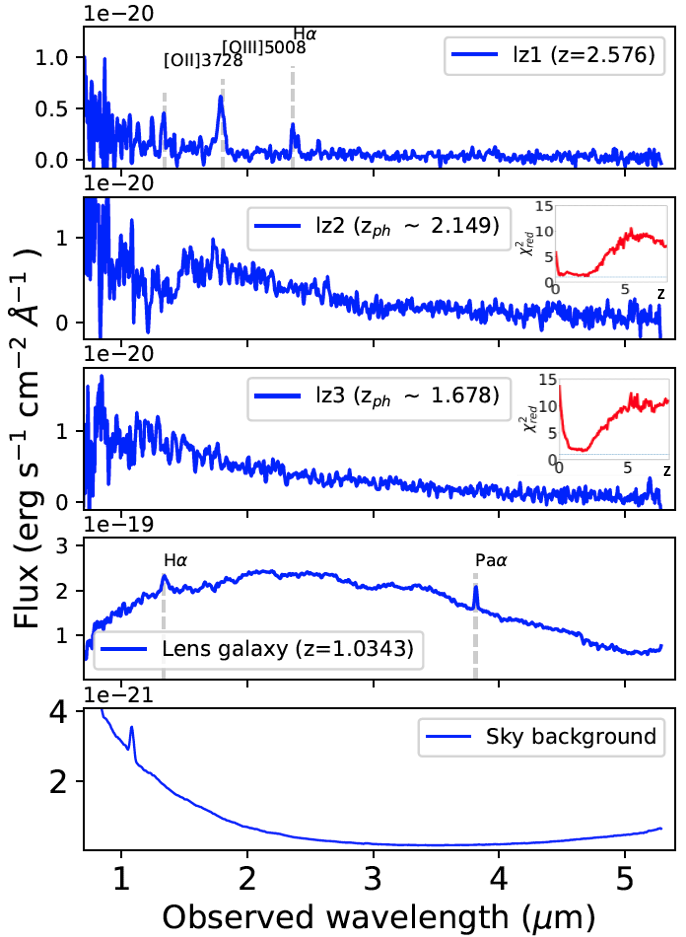}
    \caption{R100 spectra of three newly identified galaxies with z < 3 found in the FoV of SPT0311-58, together with the bright foreground (lens) galaxy. The labels indicate their redshifts (see text). For lz2 and lz3 the insets show the dependence of residuals (reduced $\chi$$^2$) as a function of the redshift, for the fits performed with EAZY (\citealp{Brammer2008}). The horizontal blue line in the inset marks $\chi$$^2_{red}$ = 1.  The sky background spectrum inferred from $\sim$ 1550 spaxels free from galaxy emission is also shown in the bottom panel.} 
    \label{fig:R100low-z}
\end{figure}

The present data allowed us to accurately determine the redshift of the foreground lens galaxy (i.e. z= 1.0343 $\pm$ 0.0002) via the identification of Pa$\alpha$ in the R2700 spectrum, improving on previous photometric determinations (z$_{phot}$=1.43$_{-0.35}^{+0.42}$, \citealp{Marrone2018}). H$\alpha$ and Pa$\alpha$ are also clearly visible in the R100 spectrum shown in Fig. \ref{fig:R100low-z}.  Using this new redshift, together with a panchromatic image (i.e. rest-frame $\sim$ 0.49 - 2.45 $\mu$m) obtained from the R100 datacube, in Appendix \ref{appendix:lens_model}, we present the magnification map produced by the lens galaxy.    

Fig. \ref{fig:R100low-z} also presents the R100 spectra of lz1, lz2, and lz3. lz1 shows very prominent emission lines, which are consistent with H$\alpha$, [OIII]$\lambda$4959,5007 and [OII]$\lambda$3726,3728 for a galaxy at z=2.576 $\pm$ 0.002. lz2 and lz3 have no emission lines, so their redshift determination is less straightforward. For them we infer the redshifts by fitting the R100 spectra with EAZY (\citealp{Brammer2008}), and find  z$_{ph}$(lz2)= 2.149$_{-0.074}^{+0.175}$, and z$_{ph}$(lz3)= 
1.678$_{-0.031}^{+0.032}$ (see also insets in Fig. \ref{fig:R100low-z}). Therefore, together with the large foreground galaxy, lz1, lz2 and lz3 do not form part of the high-z SPT0311-58 system, and they will not be discussed further in the remainder of the paper.  

\subsubsection{Spectra of the nine sources at z \texorpdfstring{$\sim$}{=} 6.9 identified with R100 spectral imaging}
\label{subsubsec:spectra_high-z }

\begin{figure*}[h]
    \centering
    \includegraphics[width=16.cm,trim= 0 0 0 0,clip]
    {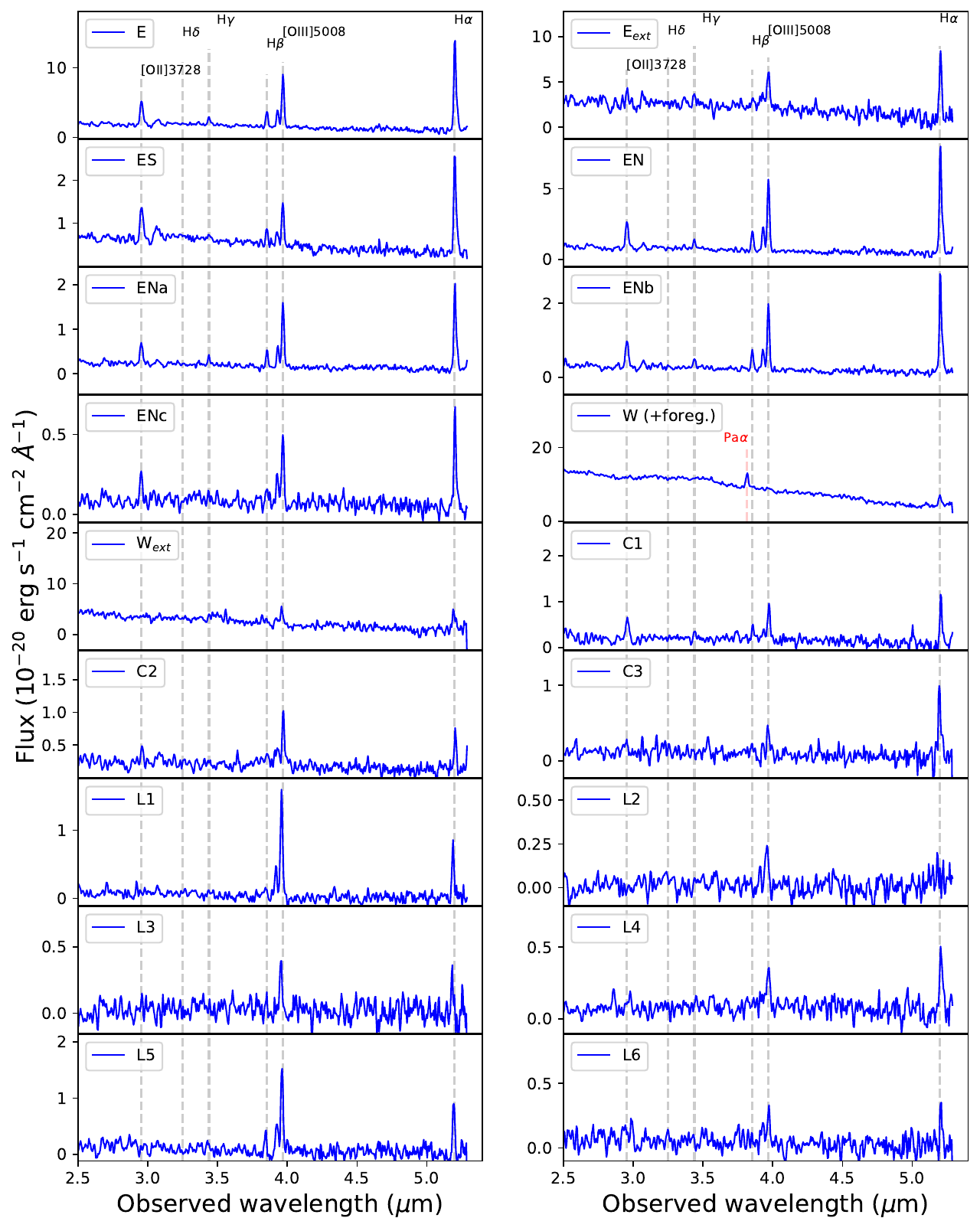}
    \caption{R100 spectra for the different z $\sim$ 6.9 sources and regions identified in the spectral images presented in Figure 2. The apertures used are shown in Fig. \ref{fig:schematic_apertures.png}, and further details in Appendix \ref{appendix:spectral_methods}. The W spectrum is contaminated by the foreground lens galaxy continuum and Pa$\alpha$ line (marked in red), which is close to \Hb\ of W (see the lens spectrum in Fig. \ref{fig:R100low-z}).The black vertical dashed lines mark the expected position of the corresponding lines for a z$\sim$ 6.92.
    }
    \label{fig:R100_spectra}
\end{figure*}

The R100 and R2700 integrated spectra for the high-z sources are presented in Fig. \ref{fig:R100_spectra} and Fig. \ref{fig:R2700spectra}, respectively. These spectra show several emission lines confirming that these sources are undoubtedly at redshifts around 6.90 (for the specific redshift values see Table \ref{table:kinematic properties}). In particular, most of the spectra include H$\alpha$, [NII]$\lambda$$\lambda$6548,6584, [OIII]$\lambda$$\lambda$4959,5008, and H$\beta$. [OII]$\lambda$$\lambda$3726,3728 (unresolved) is also present in the spectra of the brightest regions of E. 

For the W galaxy, the extraction aperture overlaps with the foreground lens galaxy (see Appendix \ref{subsec:extractionapertures}), and therefore the resulting integrated spectrum includes emission from this bright source (i.e. continuum, and Pa$\alpha$ and H$\alpha$ redshifted for z $\sim$ 1.0343). Still, this spectrum clearly shows H$\alpha$ at a redshift of $\sim$ 6.90 (Fig. \ref{fig:R100_spectra}). The line profiles for W and E in the R2700 spectra are largely oversampled due to the line broadening induced by the wide range in velocities covered by the aperture, and therefore their S/N is lower than for the R100 spectra. For these galaxies, we also obtained the spectra for their external extended emission using the E$_{ext}$ and W$_{ext}$ apertures defined in Appendix \ref{appendix:spectral_methods}.

Aperture details and line fluxes derived from these spectra are given in Tables \ref{tab:aperture_corrections} and \ref{table:line_fluxes}. Further details about the methods followed can be found in Appendix \ref{appendix:spectral_methods}.

\begin{figure}[h]
    \centering
    \includegraphics[width=0.49\textwidth]{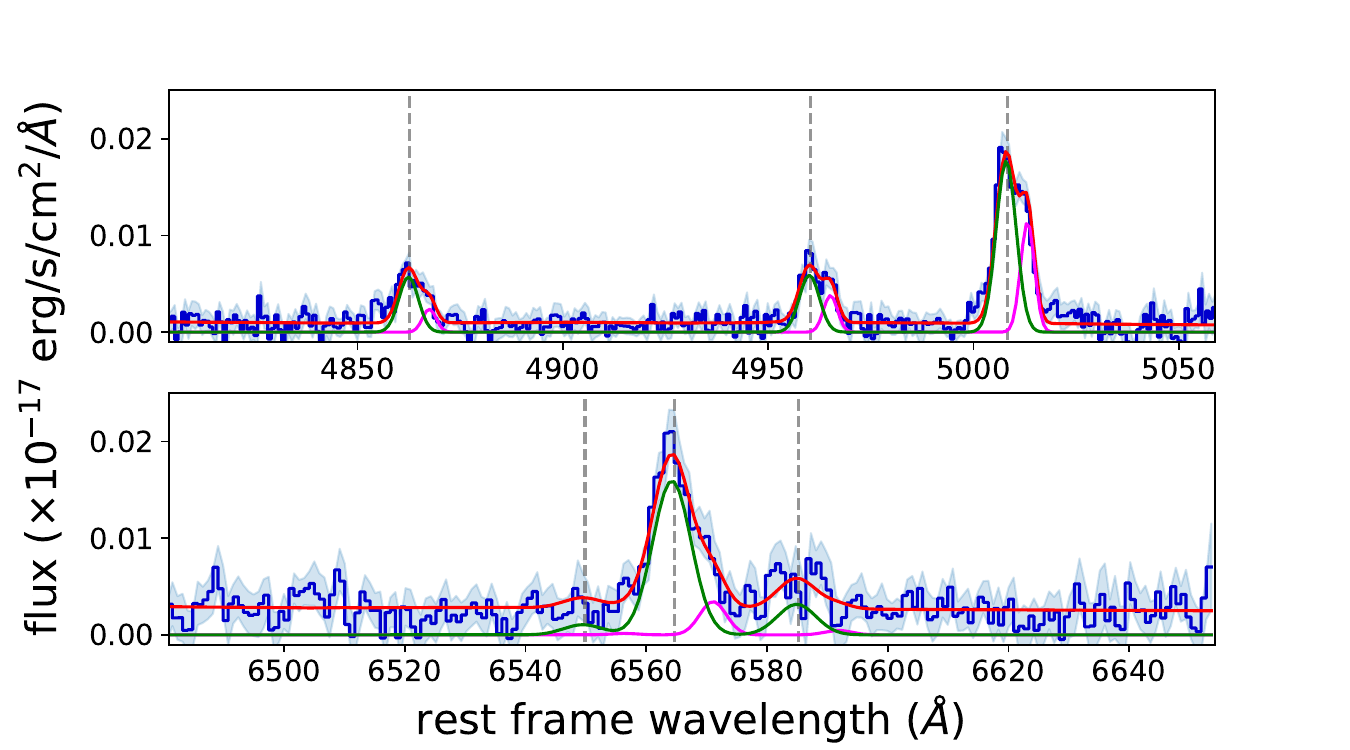}
    \caption{R2700 spectrum showing the presence of two kinematically distinct ionised gas components associated with the E galaxy, and the fainter and redshifted L7 source (see text). The panels represent the \Hb-[OIII] (top) and \Ha-[NII] (bottom) spectral regions.  The line profiles are modelled with two Gaussians, distinguished with green and magenta lines for E and L7, respectively. The total model profile is displayed in red.}  
    \label{fig:doublepeakspectrum.png}
\end{figure}

\begin{figure}[h]
    \centering
    \includegraphics[width=0.49\textwidth]{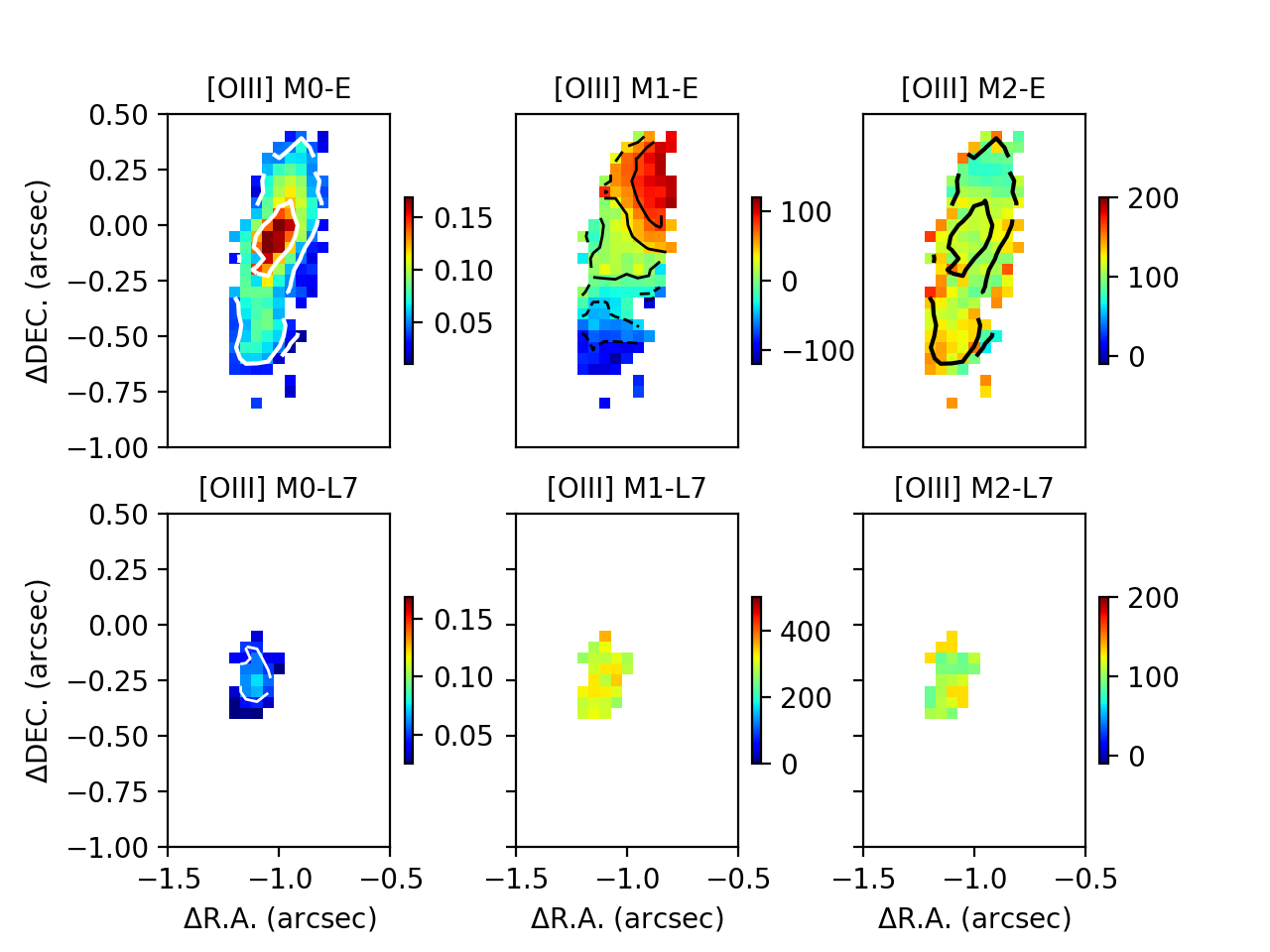}
    \caption{2D kinematic decomposition of the E and L7 galaxies. Moments 0, 1, and 2 maps for the E galaxy (upper) and the L7 source (bottom), after the kinematic decomposition (see Sect. \ref{subsubsection:Decomposition_twocomponentmaps}). From left to right, [OIII] flux map (in units of 10$^{-18}$ erg s$^{-1}$ cm$^{-2}$), velocity field (\kms), and intrinsic velocity dispersion ($\sigma_0$) map (\kms).} 
    \label{fig:Kinematics_twocomponentmaps.png}
\end{figure}

\subsubsection{Identification of L7 at z \texorpdfstring{$\sim$}{=} 6.9 via 2D kinematic decomposition}
\label{subsubsection:Decomposition_twocomponentmaps}

In addition to the nine z $\sim$ 6.9 sources identified with R100 spectral imaging and confirmed spectroscopically above, we find evidence for a tenth source overlapping with the main structure of the E galaxy. In fact, some spectra from individual spaxels close to the central region of the E galaxy show evidence for a secondary gaseous component on top of the systemic one (see Fig. \ref{fig:doublepeakspectrum.png}). Hence we performed a two-Gaussian model decomposition of these two kinematically distinct components by comparing the Bayesian information criterion (BIC) of fits performed with one and two components (\citealp{Perna2022Fits}). We also considered the possible presence of a third kinematic component that could be associated with, for example
an outflow, but the BIC analysis did not support this possibility. Figure \ref{fig:Kinematics_twocomponentmaps.png} presents the corresponding moment 0, 1, and 2 maps for the two components for the [OIII] line.

The secondary component (i.e. L7 hereafter) is relatively faint and compact and is located in projection about 0.15 arcsec (i.e. 0.8 kpc) to the southeast from the peak of the main systemic component (i.e. E). Its radial velocity is offset by $\sim$ 300 \kms with respect to E, hence allowing us a secure kinematic decomposition between the two sources. The fact the L7 is redshifted with respect to E excludes the possibility that the detected emission is due to an outflow. As we show in Sect. \ref{subsec:Ha SFR} and \ref{subsubsec: kinematics of individual galaxies}, the properties of L7 are similar to those of the other new sources discovered at z $\sim$ 6.9.  The kinematic properties of the E and L7 galaxy are discussed in Sec. \ref{subsec:kinematicsanddynamics}. 

\subsection{Summary of sources in the FoV of SPT0311-58}

We list below the various sources found in the field of view:  

\begin{itemize}

\item  Lens galaxy (z= 1.0343).
\item  Three low-z galaxies: lz1 (z=2.576$\pm$0.002), lz2 (z$_{ph}$=2.149), and lz3 (z$_{ph}$=1.678).
\item  Nine z$\sim$6.9 galaxies identified via R100 spectral imaging (C1, C2, C3, L1, L2, L3, L4, L5, L6).
\item  One z$\sim$6.9 galaxy identified through 2D spatially resolved spectroscopy of the R2700 spectra (L7).
\item  E galaxy, and its regions ES and EN (a, b, and c).
\item  W galaxy 

\end{itemize}

In addition, we obtained the spectra for the extended diffuse emission in the external regions of E and W. For all these sources, we present the integrated spectra in the sections above, except for the one identified via 2D spectroscopy (L7). For this object, we give the Gaussian model after fitting two kinematically distinct components to the spectrum (see Fig.  \ref{fig:doublepeakspectrum.png}).

\section{Stellar masses, ISM conditions, and star formation}\label{sec:ISMconditions}

In this section, we obtain the stellar masses and study the ISM properties for the high-z sources identified above. Details about the methods followed for the spectral analysis (e.g. extraction apertures, line fitting) are provided in Appendix \ref{appendix:spectral_methods}.    

\subsection{Stellar masses}
\label{subsec:Stellar masses}

Here we present stellar masses inferred from the integrated R100 spectra. We exclude W and L7 from this analysis as their stellar continua are contaminated by those of the lens and the E galaxies, respectively. We note that, although some of the newly discovered sources are weak or undetected in a spaxel-by-spaxel basis in the continuum images generated from the datacube, they show continuum emission in their integrated spectra, which typically are obtained after the combination of 25-50 spaxels. The observed fluxes were corrected by a wavelength-dependent aperture correction according to the empirical PSF derived by \cite{D'Eugenio_rotator_PSF}, and by magnification (Appendix \ref{appendix:lens_model}). As explained in Sect. \ref{subsec:reductions}, the errors in the flux provided by the pipeline were rescaled by the standard deviation in the continuum, typically by factors of 2-3. 

The SED fitting method is described in \citet{perez-gonzalezCEERSKeyPaper2023} and \citet{D'Eugenio_rotator_PSF}. In short, the NIRSpec R100 integrated spectra were compared to stellar population models from the \citet{bruzualStellarPopulationSynthesis2003} library, assuming a delayed exponential star formation history characterised by a timescale $\tau$ (taking values from 1~Myr to 1~Gyr in 0.1~dex steps) and age $t_0$ (ranging from 1~Myr to the age of the Universe at the redshift of the galaxy). The stellar metallicity $Z$ was left as a free parameter, allowing us to take all the discrete values provided by the   \citet{bruzualStellarPopulationSynthesis2003} library from 0.02 to  2.5 times solar. Nebular (continuum and line) emission was taken into account as described in \cite{perez-gonzalezStellarPopulationsLocal2003,perez-gonzalezStellarMassAssembly2008}. The attenuation of the stellar and nebular emission was modelled with a \citet{Calzetti2000} law, with $A_{\rm V}$ values ranging from 0 to 3 magnitudes in 0.1~mag steps. The stellar mass is obtained by scaling the mass-normalised stellar model to the spectrum. The derived values are included in column 6 of Table \ref{tab:Table_SFR}. The quoted stellar mass errors correspond to the uncertainties in the fit due to flux errors, but they do not include systematic uncertainties associated with the model assumptions, which are estimated to be about factors of  1.5-2. 

The stellar mass obtained for E (3.5$\pm$0.4 $\times$ 10$^{10}$ \Msun) is in excellent agreement with the one derived by \cite{Marrone2018} (3.5$\pm$1.5 $\times$ 10$^{10}$ \Msun). The predicted value from the fit at 10 $\mu$m (1.8 $\mu$Jy) is also in good agreement with the actual MIRI measurement by \cite{Alvarez-MarquezSPT} (1.6$\pm$0.3 $\mu$Jy). For this source we also perform a SED fit with CIGALE (\citealp{Boquien2019}) following the prescription and initial parameters described in \cite{Alvarez-MarquezSPT}. They adopt a two-component SFH and a \cite{LoFaro2017} attenuation law. We combine the MIRI F1000W and FIR/submm data presented in that work with the NIRSpec fluxes obtained in 22 pseudo-filters between 1 and 5.2 $\mu$m. The derived value for the stellar mass is 2.1$\pm$0.5 $\times$ 10$^{10}$ \Msun, in agreement with the results above, taking into account the expected systematic uncertainties.

The stellar masses for the newly discovered sources are based on the R100 NIRSpec data only, as they were undetected with MIRI (see Sect. \ref{subsubsec:comparisonMIRI_ALMA}). Nevertheless, we checked that the MIRI detection limit was in all cases consistent with the predicted flux for the source at 10 $\mu$m according to the SED model. We obtain a median stellar mass of $\sim$ 1.3 $\times$ 10$^{9}$ \Msun, with a range of 0.35 - 3.4 $\times$ 10$^{9}$ \Msun. We also note that recent work has shown that stellar masses inferred by SED fitting without MIRI photometry may be overestimated by $\sim$ 0.4 dex for sources at the high-redshift of \SPT (\citealp{Papovich2023}; \citealp{Wang2024}). Even taking into account this factor, our results indicate that these galaxies have already assembled a significant amount of stellar mass.  

As commented above, we do not derive the stellar mass for the W galaxy, since its flux at near-infrared wavelengths is contaminated by the foreground lens galaxy. Therefore, we adopt the stellar mass recently derived by \cite{Alvarez-MarquezSPT}, which is based on a CIGALE SED fit using HST/WFC3, JWST/MIRI, HERSCHEL/SPIRE, and ALMA photometry and the Pa$\alpha$ emission line flux. They find a stellar mass of (8.0 $\pm$ 2.4) $\times$ 10$^{10}$ \Msun\ assuming a constant SFH, and report that a similar result is obtained when adopting a two-component SFH (for further details, see \citealp{Alvarez-MarquezSPT}). Further study of the stellar population properties in \SPT is beyond the scope of the present work.

\begin{table*}
\caption{ISM properties, star formation, and stellar masses for sources in \SPT}\label{table:SFRs}    
\centering
\begin{tabular}{l c c c c c c}
\hline
\hline
ID      & H$\alpha$ flux                     &  SFR(H$\alpha$) &  $A_V$   & SFR(H$\alpha$) (dust-corr.)  &  12+log(O/H) & \Mstar \\
        & (10$^{-18}$ ergs$^{-1}$cm$^{-2}$) & (\sfr) &  (mag.) &   (\sfr)  &     & ($\times10^9$ \Msun)   \\
(1)     &                   (2)             &       (3)        &   (4)     & (5) &       (6)         &   (7)  \\     
\hline 
E       &18.8 $\pm$ 0.8 &   62. $\pm$ 2. &    1.7 $\pm$ 0.3     & 231. $\pm$ 10.     & 8.37  $\pm$ 0.06      & 35.4 $\pm$ 4.3   \\
        &               &                &    2.4 $\pm$ 0.3$^1$ & 377. $\pm$ 20.$^2$         &                       &       21.0 $\pm$ 5.2 $^5$ \\
E$_{ext}$&  11 $\pm$ 2. &   32.  $\pm$ 3.&         (0.85)       &   (61.)$^7$         &                       &                 \\
W       & 4.5 $\pm$ 0.9 &    8.  $\pm$ 1. &    5.8 $\pm$ 0.6$^1$ & 603. $\pm$ 121$^2$         &               &   (80. $\pm$24.)$^6$            \\
W$_{ext}$& 5.3 $\pm$ 1.4 &   9.  $\pm$ 1. &         (2.00)       &   (39.)$^7$                 &           &             \\
\hline
C1          & 1.5 $\pm$ 0.3 &  5.5 $\pm$ 1. &     0.7 $\pm$ 0.6    & 9.2 $\pm$ 1.8       & 8.37 $\pm$ 0.10       & 1.60 $\pm$ 0.23 \\
C2      & 0.92 $\pm$ 0.25 &  2. $\pm$ 0.4 &   1.0 $\pm$ 0.9        & 4.2 $\pm$ 1.1       & 7.97 $\pm$ 0.20       & 2.88 $\pm$ 0.68 \\
C3$^3$& 1.3 $\pm$ 0.4 &  6.2 $\pm$ 1.2 &     2.2 $\pm$ 1.6     & 33. $\pm$ 10.     & 8.56 $\pm$ 0.10       & 3.41 $\pm$ 0.94 \\
L1      & 1.2 $\pm$ 0.3   &  2.9 $\pm$ 0.4 &     1.9 $\pm$ 0.8     & 12. $\pm$ 3.        & 7.86 $\pm$ 0.16       & 0.35 $\pm$ 0.04 \\
L2      &                 &                &                       &                 &                  & 0.51 $\pm$ 0.07 \\
L3$^3$ & 0.47 $\pm$ 0.25 &1.2 $\pm$ 0.4 &     (0.85)           &   (2.3)$^7$     &                  & 1.48 $\pm$ 0.35 \\
L4      & 0.64 $\pm$ 0.22 &1.1 $\pm$ 0.3 &     (0.85)              &   (2.1)$^7$     &                  & 1.26 $\pm$ 0.19 \\
L5$^3$& 1.6 $\pm$ 0.4 &  3.9 $\pm$ 0.6 &    0.8 $\pm$ 0.6      &   7.0 $\pm$ 1.7 & 7.79 $\pm$ 0.35   & 1.17 $\pm$ 0.18\\
L6      & 0.49 $\pm$ 0.18 &1.2 $\pm$ 0.3 &     (0.85)              &   (2.0)$^7$     &                   & 0.56 $\pm$ 0.12 \\
L7$^4$& 0.85 $\pm$ 0.20 & 1.1 $\pm$ 0.3 &  (0.85)              &   (2.0)$^7$     & (7.58 $\pm$0.46)  &                 \\
\hline
\hline

\end{tabular}
\tablefoot{Column 1: Source identification. 
Column 2: Observed H$\alpha$ flux obtained by averaging values from the R2700 and R100 spectra, if S/N $>$ 3. The R100 fluxes were reduced by 10\% to account for a known calibration factor (see Appendix \ref{appendix:spectral_methods}). For E$_{ext}$, W and W$_{ext}$, only R100 data were used due to the low S/N of their R2700 spectra. In addition, for E$_{ext}$ the fits could not resolve the H$\alpha$ and the [NII] lines, and thus an average factor of 0.77 was applied to account for the [NII] emission when deriving the corresponding H$\alpha$ flux. Further details are provided in Appendix \ref{appendix:spectral_methods}. 
Column 3: SFRs obtained with the relation of \cite{kennicutt_jr_star_2012}. No dust correction is applied. For galaxies C2 and L1-L5 values have been corrected by a factor 0.5 to account for their metallicity (see text).
Column 4: Visual attenuation inferred from the Balmer decrement (\ha/H$\beta$). For E and W the values inferred from Pa$\alpha$/\ha\ are also given. For C1 and C2, $A_V$ was obtained from the R100 spectra only, as the H$\beta$ R2700 fluxes have significantly lower S/N. 
Column 5: SFR after correction for the visual attenuation in column 4.  The SFRs in columns 3 and 5 are corrected for the lensing magnification factors (see Appendix \ref{appendix:lens_model}) and the finite apertures used (Appendix \ref{appendix:spectral_methods}). 
Column 6: Metallicities based on the \cite{Curti17} calibration (see Sect. \ref{subsec:Metal}).  For L7, the metallicity is not well constrained.  
Column 7: Stellar masses (see Sect. \ref{subsec:Stellar masses}). They take into account aperture and magnification corrections. The uncertainties correspond to 16- and 84- percentiles taking into account the flux errors, but they do not include systematic uncertainties due to the model assumptions.
$^1$ $A_V$(Pa$\alpha$/H$\alpha$) using an aperture of 1$''$, as in \cite{Alvarez-MarquezSPT}. 
$^2$ Corrected with $A_V$(Pa$\alpha$/H$\alpha$).
$^3$ These sources are close the edge of the FoV, and subject to larger calibration uncertainties. 
$^4$ For this galaxy the values are derived from the 2D kinematic decomposition (see text in Sect. \ref{subsubsection:Decomposition_twocomponentmaps}).
$^5$ Stellar mass inferred with CIGALE (see text).
$^6$ Stellar mass derived by \cite{Alvarez-MarquezSPT}
$^7$ Taking an average Av (see text). 
}
\label{tab:Table_SFR}
\end{table*}

\label{Sec:ISM conditions}
\subsection{Nebular dust attenuation from the recombination lines}
\label{subsec:Attenuation}

The NIRSpec data presented here include the H$\alpha$ and H$\beta$ lines and therefore they allow us to infer the nebular attenuation due to dust from the commonly used Balmer decrement (i.e. H$\alpha$/H$\beta$). In addition, for the E and the W galaxies, the dust attenuation can also be obtained by combining the H$\alpha$ fluxes measured with NIRSpec with those for Pa$\alpha$ recently obtained with MIRI/MRS \citep{Alvarez-MarquezSPT}. To convert recombination line flux ratios into visual attenuation values (i.e. $A_V$) we take the intrinsic (dust-free) line ratio (H$\alpha$/H$\beta$)$_{int}$ = 2.79 given by \cite{Reddy23}, which is based on photoionisation modeling with CLOUDY version 17.02 (\citealp{Ferland2017}) for Case B recombination (T$_e$=10000 K, n$_e$= 100 cm$^{-3}$). This intrinsic line ratio leads to $A_V$ values that are $\sim$ 0.086 mag higher than if (H$\alpha$/H$\beta$)$_{int}$ = 2.86 (\citealp{Osterbrock1989}) is adopted. We also use the \cite{Calzetti2000} law as, e.g. in \cite{Dominguez2013}. For comparison, adopting the \cite{Cardelli1989} reddening law, the derived $A_V$ values would be 10\% lower (\citealp{RodriguezdelPino2023arXiv}). The $A_V$ values for the various sources in \SPT are presented in Table \ref{table:SFRs}.

For the E galaxy, we find that the value inferred from H$\alpha$/H$\beta$ (i.e. $A_V$(H$\alpha$/H$\beta$) =  1.7 $\pm$ 0.3) is lower than the one obtained from  Pa$\alpha$/H$\alpha$ ($A_V$(Pa$\alpha$/H$\alpha$) = 2.4 $\pm$ 0.3)\footnote{\cite{Alvarez-MarquezSPT} find a lower limit for $A_V$(Pa$\alpha$/H$\alpha$) > 4.3 mag, which is conflicting with the present results. This discrepancy can be explained by the inconsistency in the upper limit for the H$\alpha$ found with MIRI and the NIRSpec measurement. }.  In local dusty objects such as luminous and ultraluminous infrared galaxies (i.e. LIRGs and ULIRGs, respectively), several authors have also reported that the visual attenuation derived from Balmer lines ratios is lower than that obtained from ratios involving infrared lines (e.g. \citealp{Piqueras-Lopez2013extinction}; \citealp{Gimenez-Arteaga2022}). Moreover, for intermediate redshifts (i.e. 1 $<$ z $<$ 3.1) \citet{Reddy23} have found that ratios involving Paschen lines lead to reddening measurements that are often larger than those obtained with the Balmer decrement. These results are generally explained by assuming that part of the flux in the infrared lines comes from regions that are optically thick for the Balmer lines. In other words, due to the lower obscuration at longer wavelengths,  H$\beta$, H$\alpha$ and Pa$\alpha$ trace increasingly deeper dusty regions, which leads to the above-mentioned differences.

For the W galaxy, we cannot reliably infer $A_V$(H$\alpha$/H$\beta$), as in the R100 spectrum H$\beta$ is blended with Pa$\alpha$ from the lens galaxy, and in the R2700 spectrum the \Ha\ and \Hb\ fluxes are noisy due to the broadening induced by the large range in velocities covered by the aperture.
However, the H$\alpha$ and  Pa$\alpha$ fluxes obtained with NIRSpec (R100) and MIRI, respectively, imply $A_V$(Pa$\alpha$/H$\alpha$) = 5.8 $\pm$ 0.6. 
This value confirms that W is significantly more attenuated by dust than E, as previous studies (e.g. \citealp{Marrone2018}) have already shown, based mainly on far infrared measurements (see ALMA continuum map at 160 $\mu$m in Fig. \ref{fig:R100imaging}). 

For several of the newly discovered z$\sim$ 6.9 sources (i.e. C1-3, L1, L5) the $A_V$(H$\alpha$/H$\beta$) could be estimated. The obtained values range from $\sim$ 0.7 to 2.2 mag, and indicate that in general they have lower attenuation than E and W. We note that the uncertainties for the individual values are large, and in some cases they are also consistent with low or no attenuation. The lack of ALMA continuum emission in these sources also supports a low dust content. The regions in the eastern part (C1, C2, L5) have, on average, a lower attenuation (<$A_V$> $\sim$ 0.85 mag) than the ones in the western region (C3, L1; <$A_V$> $\sim$ 2 mag). We used these average values to estimate the attenuation for sources that do not have reliable H$\beta$ fluxes, including the extended regions W$_{ext}$ and E$_{ext}$  (see Table \ref{table:SFRs}).

\begin{figure*}[h]
    \centering
    \includegraphics[width=17.cm,trim= 0 0 0 0,clip]
    {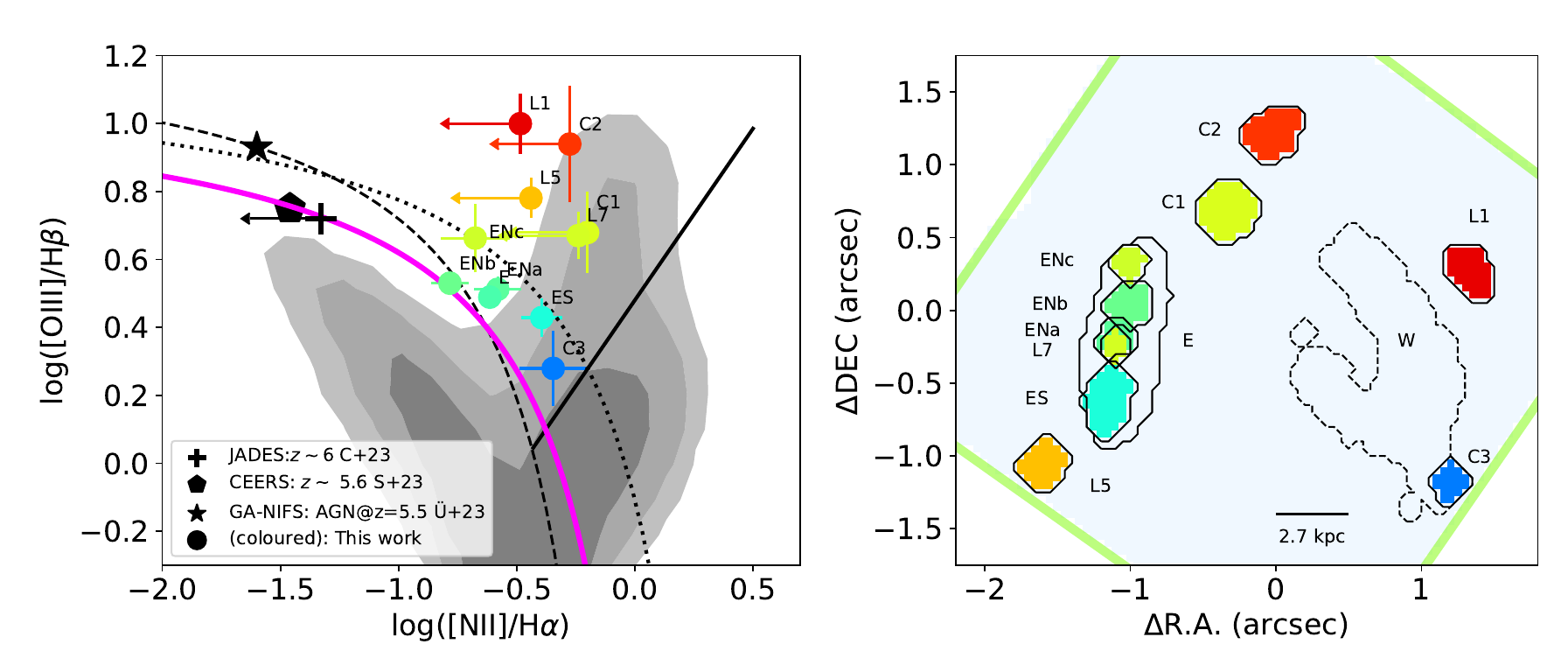}
    \caption{Classical N2-BPT diagram and ionization map for the z $\sim$ 6.9 sources in SPT0311-58. Left: BPT diagram 
    colour coded according to the [OIII]$\lambda$5007/H$\beta$ value. The line ratios are derived from the integrated R2700 spectra, and include the subregions within E and the five external z $\sim$ 6.9 sources with reliable line fluxes. Upper limits correspond to 3$\sigma$. Grey-shaded regions show the zones of the diagram covered by 70, 80, and 90 percent of the local SDSS sample of star forming galaxies and AGN. (\citealp{York2000}, \citealp{Abazajian2009}).  
    Below the dashed line, local galaxies are ionised by SF (\citealp{Kauffmann2003}), while above the dotted line the presence of AGN is required (\citealp{Kewley2001}). The region beneath the black solid line corresponds to LI(N)ERs (\citealp{Cid-Fernandes}). The magenta line is the locus of z $\sim$ 2--3 star-forming galaxies found by \cite{Strom2017}. The black symbols correspond to other galaxies at z > 5 observed with JWST/NIRSpec (C+23: \citealp{Cameron2023}; S+23: \citealp{Sanders2023}; Ü+23: \citealp{Ubler23}). Right: Spatial distribution of the different galaxies in the SPT0311-58 system. The apertures used (Appendix \ref{fig:Apertures}) are colour coded according to their [OIII]$\lambda$5007/H$\beta$ values, as in the left panel.    
    } 
    \label{fig:Diagnostic_NII}
\end{figure*}

\subsection{Emission line ratio diagnostic diagrams}
\label{subsubsec:BPT}

We now discuss the excitation conditions in the different high-z galaxies identified in \SPT.  In Fig. \ref{fig:Diagnostic_NII} (left) we present the classical $N2$ $\equiv$ [NII]/H$\alpha$ vs. $R3$ $\equiv$ [OIII]/H$\beta$ BPT diagram (hereafter N2-BPT; \citealt{Baldwin}). We include the z $\sim$ 6.9 sources with reliable line flux ratios (i.e. C1, C2, C3, L1, L5, L7), in addition to the subregions within E. The line fluxes have S/N $>$ 3, except for the [NII] flux of some sources for which upper limits are provided. Colours are assigned  according to the [OIII]/H$\beta$ value. The figure shows the zones of the diagram covered by the z $\sim$ 0 SDSS galaxies (\citealp{Abazajian2009}) (shaded grey), together with several well-established lines that demarcate the different excitation regimes at low z. We also include in the plot (magenta line) the locus of the intermediate redshift (i.e. z $\sim$ 2--3) star-forming galaxies from the KBSS-MOSFIRE survey (\citealp{Strom2017}), and results at redshifts higher than 5 recently obtained with JWST (\citealp{Cameron2023}, \citealp{Sanders2023}, \citealp{Ubler23}).  

\begin{figure}[h]
    \centering
    \includegraphics[width=0.5\textwidth]
    {Figures/O32R23_v2.pdf}
    \caption{$R23$ -- $O32$ diagnostic diagram. The blue dots represent the galaxies in \SPT and are based on  R2700 fluxes (except when the line was only detected in R100). They are corrected for attenuation and aperture. Grey-shaded regions show the zones covered by 80, 90, and 99 percent of the local SDSS sample of star forming galaxies (\citealp{York2000}, \citealp{Abazajian2009}). Green and red isocontours correspond to mass ranges log(\Mstar/M$_\odot$) = [8.5 - 9.5] and [9.5 - 10.5], respectively. The black symbols are for stacked spectra of other high-z samples observed with JWST/NIRSpec, as indicated by the labels (C+23: \citealp{Cameron2023}; S+23: \citealp{Sanders2023}).} 
    \label{fig:BPT-O32_R23.png}
\end{figure}


Fig. \ref{fig:Diagnostic_NII} shows that the sources in \SPT occupy zones of the N2-BPT diagram barely populated by local galaxies, and are distributed along the SF/AGN demarcation lines at z=0, covering more than 0.7 dex in $R3$. Although most of the sources occupy the composite/AGN zones of local galaxies, they do not exhibit clear evidence of AGN activity (e.g. Balmer broad lines or signs of strong outflows). However, we do not exclude the presence of AGN, as at high-z their associated outflows do not appear to be very prominent (\citealp{Maiolino2023_bh}, \citealp{Carniani2023_outflows}, Scholtz et al., in prep.).  These z $\sim$ 6.9 sources are in a region shifted by 0.1-0.3 dex (in $N2$ or $R3$) with respect to the star-formation locus at z$\sim$ 2-3,  enlarging further the offset with respect to the z=0 population. We note that none of the galaxies are in the local LI(N)ER part of the diagram, despite the likely presence of shocks according to the observed kinematics (see Sect. \ref{subsec:kinematicsanddynamics}, and  \citealp{Spilker2022}). The sources in \SPT with the highest $R3$ (i.e. C2, L1, and L5) are in similar regions of the N2-BPT diagram as other galaxies at z $>$ 6 recently observed with JWST (e.g. \citealt{Cameron2023}, \citealt{Sanders2023}), and also have faint or undetected [NII]$\lambda$6583 fluxes.  

The spatial distribution of $R3$ reveals interesting trends (Fig. \ref{fig:Diagnostic_NII}, right panel). First, the sources with the lowest $R3$ (ES, C3) are found within or close the massive galaxies E and W. Second, E shows a clear variation in $R3$ along its major axis, with increasing values from south to north (i.e. ES, ENa, ENb, ENc). We note that L7 is seen in projection over the main structure of E, but it is a separated source (see Sect. \ref{subsubsection:Decomposition_twocomponentmaps}).  Third, the sources with the highest $R3$ (C2, L1, L5) are found in the external regions. 
  
We now compare the observed positions in the N2-BPT diagram with the predictions of the theoretical study by \cite{Hirschmann22} based on the coupling of the state-of-the-art cosmological IllustrisTNG simulations (\citealp{Springel2010IllustrisTNG}) with new-generation nebular-emission models (\citealp{Hirschmann2019}). 
While sources in \SPT show a wide range in $R3$ (i.e. 0.25 -- 1), TNG50 galaxies at z $\sim$ 5-7 have $R3$ > 0.60   (\citealt{Hirschmann22}, their Fig.6). The external sources (i.e. C1, C2, L1, L5) have observed values within this predicted range. However, for regions within or close the two main galaxies (i.e. ES, ENa, ENb, ENc, C3) the observed $R3$ are < 0.60, and only agree with the theoretical predictions at low or intermediate redshifts (z$\sim$ 0.5--3). Therefore, these galaxies seem to have exceptional properties for their z, according to theoretical expectations.

In Fig. \ref{fig:BPT-O32_R23.png}, we present the galaxies in the dust-corrected $O32$-$R23$ diagram.  
$R23$ $\equiv$ log(([OIII]$\lambda$$\lambda$4959,5008 + [OII]$\lambda$$\lambda$3726,3729)/\hb) is a proxy of the total excitation, while $O32$ $\equiv$ log([OIII]$\lambda$5008/[OII]$\lambda$$\lambda$3726,3729) is sensitive to the ionisation parameter and the metallicity (\citealp{Kewley2019ARA&A}, \citealt{Cameron2023}). This diagram reflects again the different ISM conditions of the sources in \SPT in comparison with those commonly found in local galaxies. The large range in $O32$ indicates a significant variation of the ionisation parameter and/or the metallicity among the sources in the FoV.  It also shows the dichotomy between the external regions (C2, L1, L5) and those associated with the main structures of the E and W galaxies (C3, ES, ENa, ENb, ENc), with the latter group having lower $O32$.  To take into account the interdependence of stellar mass and metallicity, which are defined by the mass-metallicity relation (MZR; e.g. \citealp{Curti2023JADES-MZR}), in Fig. \ref{fig:BPT-O32_R23.png} we distinguish two mass ranges for the local sample: log(\Mstar/M$_\odot$) = [8.5 - 9.5] and [9.5 - 10.5]. The former is in principle more appropriate for comparison with the newly discovered sources, while the latter covers the mass of the E galaxy, and therefore adequate for its subregions. The figure shows that again in this diagram C3 and most of the regions within E, have positions closer to those occupied by local galaxies. The external sources C2, L1, and L5 occupy a distinct position in the diagram, displaced with respect to local galaxies with similar stellar mass. They are close to the average position of the field galaxies at z $\sim$ 5.6 in CEERS (\citealp{Sanders2023}, log(\Mstar/M$_\odot$) = 8.57), but seem to have a lower $O32$ value with respect to the less luminous galaxies in JADES (\citealp{Cameron2023}). Still, the low statistics prevent us to perform a solid comparison with these samples.  

In summary, we find that the ISM conditions in the sources of SPT0311-58 are very diverse. While some of these z $\sim$ 7 objects have properties similar to galaxies at these redshifts recently observed with JWST, others have conditions closer to those found at lower redshifts. Most occupy the region of SF/AGN composite ionisation, and none are in the local LI(N)ER region despite the likely presence of shocks. The diagnostic diagrams indicate significant differences in the ionisation parameter and/or in the metal content across the observed sources. In the next section we study in more detail the metal content in the system.

\begin{figure}[h]
    \centering
    \includegraphics[width=0.49\textwidth]{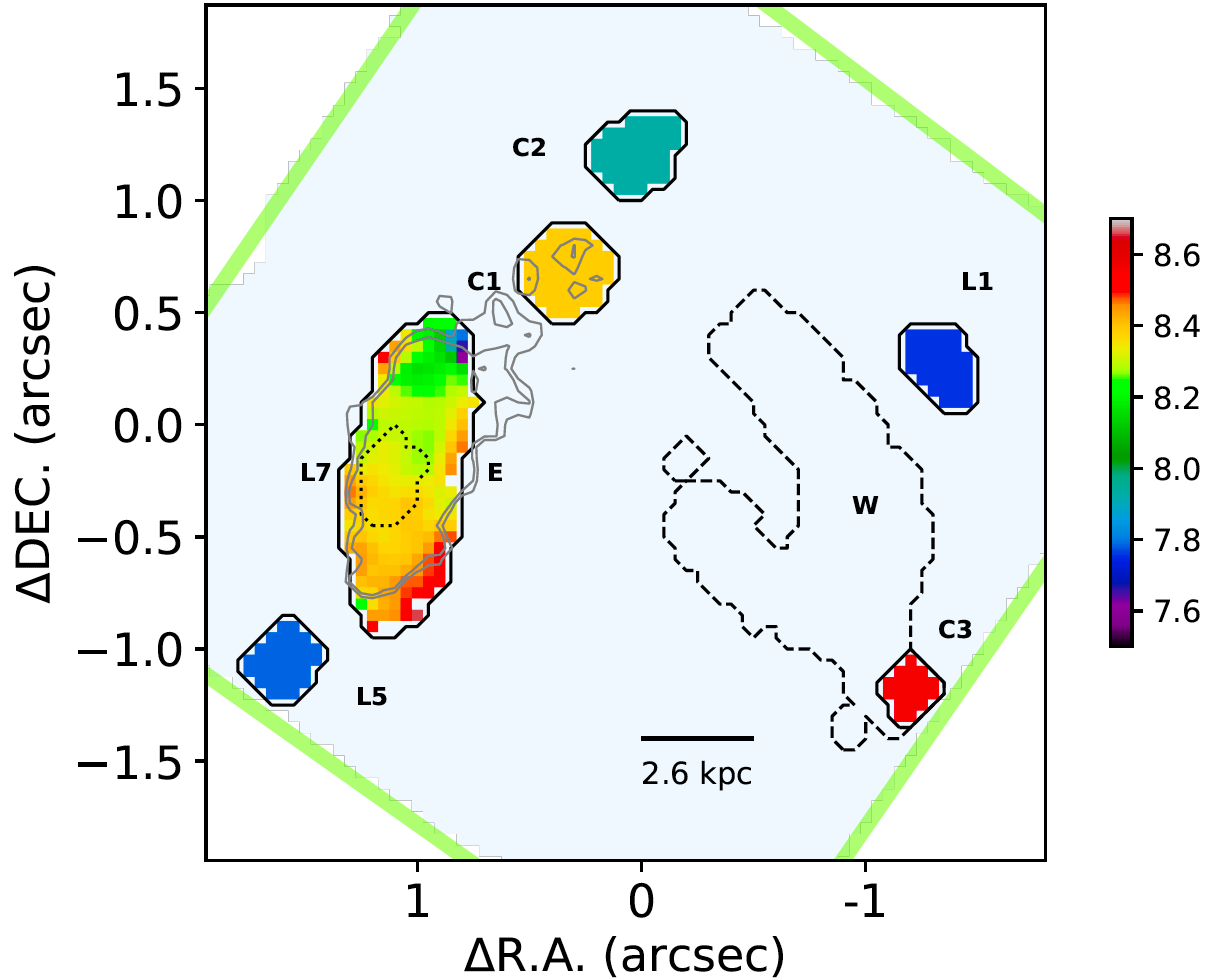}
    \caption{Metallicity map based on the calibrations of \cite{Curti17,Curti2020} in the 12+(O/H) scale. For the E galaxy, spaxel-by-spaxel information is provided. For the external regions, the value obtained from the integrated spectra is assigned to the entire aperture. 
    The region where the galaxy L7 appears in projection over E, is indicated with a dotted line. The line fluxes were corrected with the mean $A_V$ value for the apertures. The faint grey lines correspond to two low-level isocountours of the H$\alpha$ emission (Fig. \ref{fig:R100imaging} (c)) (see text).   
    } 
    \label{fig:Metal_map.png}
\end{figure}


\subsection{Metallicity}
\label{subsec:Metal}

In Fig. \ref{fig:Metal_map.png}, we present the metallicity map of SPT0311-058 according to the calibration of \cite{Curti17, Curti2020}. This calibration is based on local samples, but it has the advantage of covering well the metallicity range observed in the system, and it shows relatively low scatter. We also comment below the results when using the calibration by \cite{Sanders2023}, which is based on samples that cover well the redshift of \SPT, but presents larger scatter than local ones due to the lower number of sources available and potential differences in the ISM conditions at high-z. We followed the procedure developed by \cite{Curti2023JADES-MZR}, which explores the log(O/H) parameter space with a Markov Chain Monte Carlo (MCMC) algorithm and finds the maximum likelihood taking into account the uncertainties of the observed line ratios and the dispersion of the individual index calibrations. For the E galaxy, the metallicity is derived for each spaxel (after a 3 x 3 binning) while for the external sources, and due to their lower S/N, the integrated spectra are used to derive a global value. For E, C1, C2 and L1 the metallicity is inferred from the $R2$, $R3$, $R23$, $O32$, and {\it \text{\^R} } indexes, where  $R2$ $\equiv$ log(\OIIlines/\Hb) and {\it \text{\^R} } $\equiv$ $0.47\ \text{$R2$} + 0.88\ \text{$R3$}$, as defined by \cite{Laseter2023}. For C3, $R3$ and $N2$ were used, while for L5 and L7 only $R3$ was available due to the lack of reliable line fluxes to derive other indexes. In Table \ref{tab:Table_SFR} the individual metallicity values are presented. 

Fig. \ref{fig:Metal_map.png} shows that the metallicity spans about 0.8 dex across the FoV. On the one hand, the southern regions of E and C3 have the highest metallicities with values around 8.4-8.6 in the 12+log(O/H) scale. On the other hand, external galaxies C2, L1, and L5 have low metal content, with values in the range $\sim$ 7.6-8.1. This pattern is consistent with the mass metallicity relation (MZR, \citealp{Curti2023JADES-MZR}), according to which the most massive galaxies (and therefore their subregions) have the highest metallicities, and suggests that C3 is closely associated with the massive W galaxy. The northern region of E, which shows low metallicity and therefore differs from this general behaviour, is discussed below.   
The calibration by \citealp{Sanders2023} leads to qualitatively similar results, although it finds higher metallicity (by $\sim$ 0.2 dex) in the metal rich regions C1, and E, while leaving unaltered the values for C2 and L1, and suggesting lower metal content in L5 (by $\sim$ 0.2 dex). Therefore, this calibration indicates an even larger range in metallicty across the FoV, though it also increases the associated uncertainties of individual determinations.         

Interestingly, the E galaxy shows a prominent metallicity gradient along its major axis with a total peak-to-peak variation of $\sim$ 0.7 dex, and also local inhomogeneities at subkpc scales. The southern regions are the most metal-rich (i.e. $\sim$ 8.50), while the lowest metallicity (i.e. $\sim$ 7.8) is found to the northwest. This metal-poor region is adjacent to a zone of diffuse H$\alpha$ (see grey isocontours in Fig. \ref{fig:Metal_map.png}) and [CII] emission (Fig. \ref{fig:R100imaging} c and f), indicating the presence of ionised and neutral gas. This suggests that the accretion of low metallicity gas into the main structure of E in the northwest may be the cause for the observed gradient. Gas accretion has been proposed to explain metal gradients and inhomogeneities at high- and low-z (e.g. \citealp{Cresci2010_metalgradientsinflows},  \citealp{Kumari2017},  \citealp{Curti2020_metalgradients},    
 \citealp{RodriguezdelPino2023arXiv}). In Sect. \ref{subsubsect:dicussion_turbulent_disc}, we will discuss further this scenario, as well as other potential alternatives (i.e. merger event, outflows), also taking into account the kinematic properties of the E galaxy.   

\cite{Litke2023} have recently found a nearly solar metallicity in the massive galaxies of \SPT based on the observed far-infrared [OIII]$\lambda$ 88$\mu$m /[NII]$\lambda$ 122$\mu$m emission line ratio, and the calibration proposed by \cite{pereira-santaella_far-infrared_2017} (i.e. 0.8 $\pm$ 0.4 Z$_\sun$ for E; 1.4 $\pm$ 1.0 Z$_\sun$ and 0.9 $\pm$ 0.4 Z$_\sun$ for the northern and southern regions of W, respectively). 
To compare their results with our findings for E, we computed the global metallicity over the integrated E spectrum.  Using the $R2$, $R3$, $R23$, $O32$, and {\it $\text{\^R}$} indexes and the calibration by \cite{Curti17, Curti2020}, we infer an average value of 0.48 $\pm$ 0.07 Z$_\sun$. Although the difference with the Litke at al. result is not significant taking into account the uncertainties, it could be interpreted as due to the different regions probed by the FIR and the optical lines. While the FIR offers access to deeper, dustier, and likely more metal-rich zones, the optical lines come from the less extincted and likely more metal-poor regions. Finally, the comparison of our metallicity value for C3  (i.e. 0.75 $\pm$ 0.18 Z$_\sun$), with the results by  \cite{Litke2023} for the southern part of the W galaxy show good agreement.   

In summary, we find a wide range of metallicities (by $\sim$ 0.7 dex) across the galaxies in \SPT. The newly discovered  galaxies seem to have lower metal content than the relatively metal-rich massive E and W galaxies. The E galaxy also shows a metallicity gradient that is likely evidence for the presence of pristine gas accretion.  In Sect. \ref{subsubsect:dicussion_turbulent_disc} we discuss further this and other alternative scenarios, also considering the kinematic properties of the system.

\subsection{H$\alpha$-based star formation}

\label{subsec:Ha SFR}
The present observations allow us for the first time to measure H$\alpha$ fluxes in this system, and therefore to infer the unobscured ongoing star formation using this well-calibrated tracer. We note that \SPT is at about the highest z at which H$\alpha$-based SFR can be derived at the sensitivity and angular resolution provided by NIRSpec (i.e. for z$>$ 7, \Ha\ is redshifted outside the NIRSpec range).    

We estimate the nearly instantaneous (i.e. last 3-10 Myrs) star formation rate (SFR) from the \ha\ luminosity following the commonly used expression provided by \cite{kennicutt_jr_star_2012} (see also \citealp{Murphy2011}; \citealp{Hao2011}):  SFR(\Ha)/(\sfr) = 5.37 $\times$ 10$^{-42}$ L(\Ha)/(erg s$^{-1}$). 

This relation presumes constant SFR over the mentioned timescale, solar metallicity and a Kroupa IMF (\citealp{Kroupa2001}) \footnote {Results for a Chabrier IMF (\citealt{Chabrier2003}) are nearly identical (c.f. \citealt{kennicutt_jr_star_2012}).}, and also assumes that all ionising photons actually ionise the ISM. We note, however, that changing some of these assumptions may have significant effects. For instance, for subsolar metallicities the SFRs provided by this equation should be corrected down because metal poor massive stars are more efficient in producing ionising photons, which leads to higher L(\ha) per unit of SFR as discussed by \cite{Reddy2022} (see also \citealp{Shapley2023}). To take into account this effect, and based on the results of the previous section, the SFRs obtained with the expression quoted above for galaxies C2, L1-L5 have been corrected accordingly by a factor 0.5. For the more metal-enriched E and W galaxies (and C1 and C3), we have maintained the solar abundance assumption. This allows us a more direct comparison with previous SFR estimates for E and W (\citealp{Marrone2018}, \citealp{Alvarez-MarquezSPT}), and with determinations based on general calibrations (e.g. Total Infrared Luminosity L(TIR)-SFR,  \citealp{Calzetti2013}) that have used solar metallicity. The individual H$\alpha$ fluxes and SFRs(H$\alpha$), corrected and uncorrected for dust attenuation, are presented in Table \ref{table:SFRs}. 

As we can see in the table, and also in panel c of Fig. \ref{fig:R100imaging}, the main body of the E galaxy is the region in the FoV with the highest H$\alpha$ flux, and therefore where most of the unobscured (i.e. uncorrected by dust attenuation) star formation is taking place at a rate of $\sim$ 62~\sfr.  The dust-corrected SFR (377 \sfr) represents a large fraction (i.,e. $\sim$ 70-80 percent) of the total SFR obtained by \cite{Marrone2018} from a SED analysis which includes the far-IR (ALMA) dust continuum emission (545 $\pm$ 175 \sfr), or by using the L(TIR)-SFR calibration by \cite{Calzetti2013} ($\sim$ 466 \sfr).  In addition to the \Ha\ flux associated with E, a low surface brightness H$\alpha$ emission extends beyond its main structure, particularly in the direction of source C1.  This external \Ha\ flux is obtained integrating within aperture E$_{ext}$ (see Fig. \ref{fig:Apertures}), and has an associated unobscured SFR(\Ha) of $\sim$ 32 \sfr ($\sim$ 61 \sfr dust-corrected). 
 
For W the inferred unobscured SFR(\Ha) is $\sim$ 8 \sfr, which is a tiny fraction of the SFR obtained by \citealp{Marrone2018} from a SED analysis including the FIR emission (i.e. 2900 +/- 1800 \sfr), or  from the L(TIR)--SFR relation by \citealt{Calzetti2013} (3350 \sfr ). 
In this case, the attenuation correction factor is extremely large, as $A_V$ is 5.8 magnitudes. The dust-corrected SFR(H$\alpha$) is $\sim$ 600 \sfr, which is about one-fifth of the total SF inferred including the far-infrared dust emission. This suggests that W hosts a large fraction of star formation that is totally obscured at rest-frame optical and near-IR wavelengths. We note that a detailed comparison of the SFR obtained with different tracers should also take into account the timescale assumed for their calibration and the star formation history (SFH) of the source under study. For example, post-starburst (PSB) galaxies (where star formation terminated rapidly and recently) have systematically higher SFR(FIR)s than what is estimated from recombination lines alone. This is observed both in the local Universe (\citealp{Baron2023}) as well as at z $\sim$ 1 (\citealp{Wu2023}). However, these PSB galaxies also display lower nebular attenuation than what we infer for the W galaxy; in our case, because we are dealing with an extremely dusty galaxy, we favour the star-forming interpretation, and explain the discrepancy between the SFR(H$\alpha$) and SFR(FIR) as due to heavy dust attenuation. The most rigorous approach would be to infer the full SFH, but this is beyond the scope of the present study. 

For the newly discovered sources at z $\sim$ 6.9, the unobscured SFR(\Ha) ranges from 1 to 6 \sfr, and totalling 14 \sfr. The sources with $A_V$ measurements (i.e. C1, C2, C3, L1, L5), have individual dust-corrected SFR(\ha) ranging from $\sim$ 4 to 30 \sfr. For the other sources, the star formation is not well constrained as $A_V$ cannot be reliably derived (see Sect. \ref{subsec:Attenuation}). Using for them the average $A_V$ derived for the other sources, the combined dust-corrected SFR(\ha) for all these objects is 74 \sfr, a small amount compared with the total SFR taking place in E and W. As the new sources are undetected in ALMA continuum, their dust-corrected SFR(\Ha) likely represent the total SFR. 

In summary, we find significant unobscured ongoing star formation activity within and around the E galaxy ($\sim$ 94 \sfr). The dust-corrected SFR(\Ha)  ($\sim$ 430 \sfr) recovers a large fraction of the total SFR obtained including the FIR emission. In contrast, for W the current unobscured SFR(\Ha) is small (8 \sfr), representing a tiny fraction of the total SFR of this galaxy. An important fraction of the SFR in this galaxy is totally obscured at optical and near-IR wavelengths, and cannot be recovered with the attenuation correction. As for the newly discovered external sources, their unobscured SFR(\Ha)s are relatively modest with individual values $\sim$ 1 - 6 \sfr. Correction from dust still leaves the combined contribution to these sources to the total SFR of the system as minor (i.e. $\sim$ few percent).        

\section {Kinematic properties}
\label{subsec:kinematicsanddynamics}

This section is devoted to the kinematic properties of the different sources of \SPT. First, we present the relative motion and velocity dispersion of the newly discovered sources at z $\sim$ 6.9. We also derive their dynamical masses, and compare them with the stellar masses.  Second, we focus in more detail on the E and W galaxies, and discuss the kinematics for the warm and cold gas phases inferred from NIRSpec and ALMA, respectively.

\subsection{Newly discovered z \texorpdfstring{$\sim$}{=} 6.9 sources: Relative velocities, velocity dispersions, and dynamical masses}
\label{subsubsec: kinematics of individual galaxies}

\begin{figure*}[h]
    \centering
    \includegraphics[width=18.7cm,trim= 0 0 0 0,clip]
{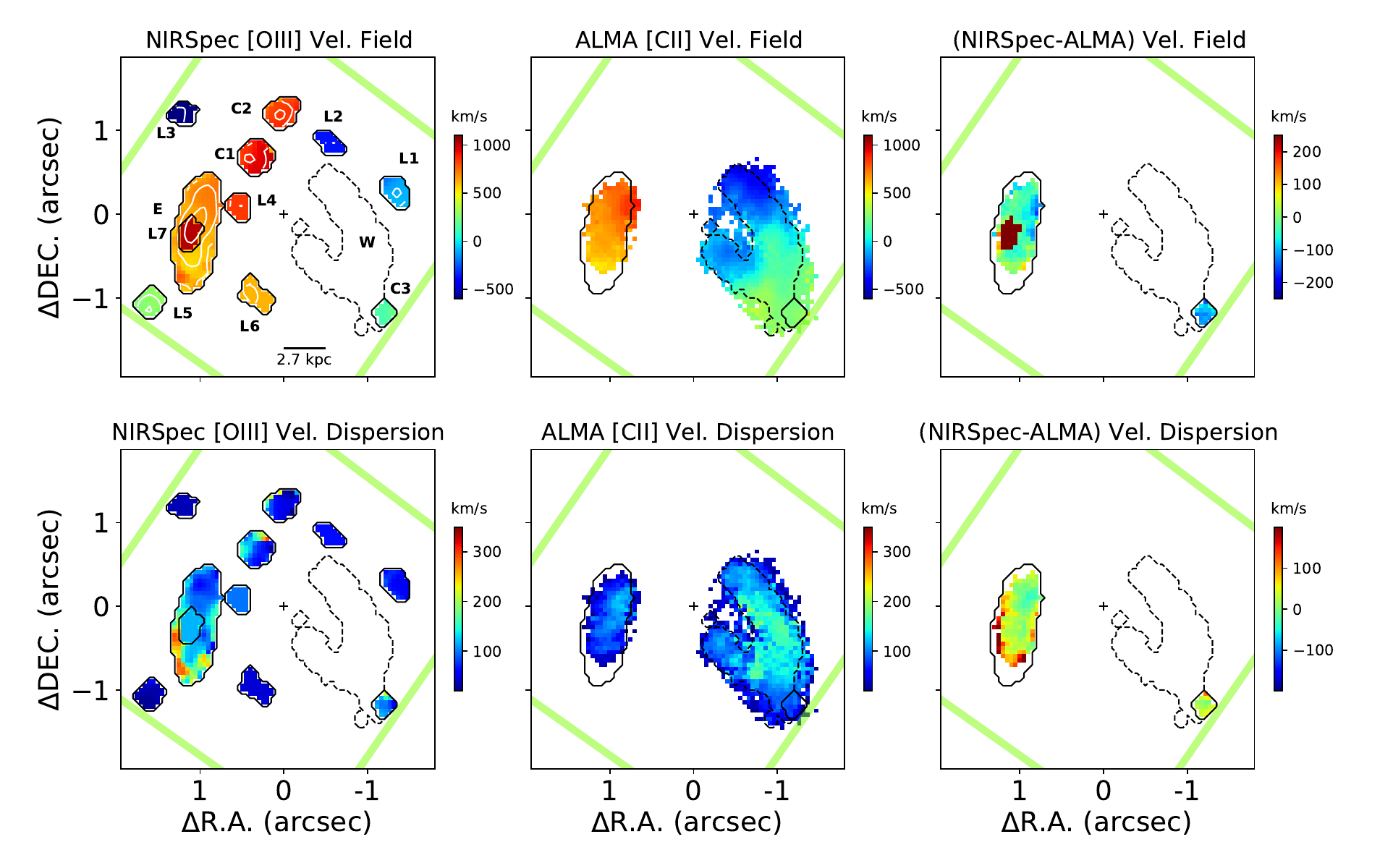}
    \caption{Kinematic maps of \SPT. Upper panels: 
    Large-scale velocity fields from the high resolution NIRSpec IFU spectra in the [OIII]- \Hb\ complex (left), from the [CII]$\lambda$158 line observed with ALMA (centre), and difference (right). Lower panels: similar maps for the intrinsic velocity dispersion ($\sigma_0$). The NIRSpec maps were obtained after a 3 $\times$ 3 spaxel binning, except for L2, L3, L4, L6 and L7 for which, due to their low S/N, the values from the integrated spectrum were assigned to the entire extraction aperture. The values for L7 were obtained after the kinematic decomposition of two overlapping systems (Sect. \ref{subsubsection:Decomposition_twocomponentmaps}). The relative velocity values are in \kms , taking as reference a redshift of 6.902. The NIRSpec velocity dispersions are corrected for the instrumental broadening according to the dispersion curves provided by \cite{Jakobsen2022}   
    } 
    \label{fig:Kinematics}
\end{figure*}

Fig. \ref{fig:Kinematics} (upper-right panel) shows the large-scale velocity field in \SPT derived from the R2700 [OIII] line. In this figure the differences in redshift are interpreted as relative velocities 
\footnote{We note that, if the observed redshifts were alternatively interpreted as spatial separations only, the proper distance between the two furthest separated galaxies (i.e, L3 and C1) would be about 3.6 Mpc. This value is similar to the rough estimate of the extent for the overdense region with submm galaxies ($\sim$ 2.6 Mpc; \citealp{Wang2021}).},  taking z = 6.902 the origin for the velocities (i.e. the redshift of the most massive W galaxy inferred from the [CII] line, Sect. \ref{subsec: Wgalaxy}). The figure shows that the ten newly discovered galaxies have a very large range of (projected) velocities: from -595 \kms\ for L3 up to +932 \kms\ for C1 (see Table \ref{table:kinematic properties} for the individual redshifts).  These velocities seem randomly distributed across the FoV, with sources with high positive and negative values relatively close in projection (e.g. L2 and C2, L3 and C1). This significant spread in velocities suggests that the mass of the halo must be very large. In Sect. \ref{Sec:Discussion} we discuss further the relative motions of the sources within the system.

The integrated velocity dispersion ($\sigma$, see Table \ref{table:kinematic properties}) can be used to estimate the dynamical mass (M$_{\rm dyn}$) of the galaxies through the formula:  
\begin{equation}\label{eq:mdyn}
    M_{\rm dyn} = K \frac{\sigma^2 R_e}{G},
\end{equation}
where $R_e$ is the effective radius, $G$ is the gravitational constant, and $K$ is a factor that depends on the mass distribution in the galaxy. We adopt $K$=5$\pm$0.1, following the best-fitting virial relation calibrated by \cite{Cappellari2006}. We also applied the correction suggested by \cite{Ubler23}, to account for the differences in the integrated velocity dispersion of the ionised gas and the stars found by \cite{Bezanson2018}. For the present range of integrated $\sigma$, this $\Delta$ log($\sigma$/(\kms)) correction varies from 0. to +0.18. 
For the determination of the size, we fit the emission in each source with a 2D Gaussian, and estimate $R_e$ from the largest of the two standard deviations provided by the fit, after deconvolving with the PSF. To derive the properties of the PSF, we used the data cube of the standard star (Sect. \ref{subsec:ancillary}). We first try to infer the sizes from a panchromatic image generated from the R100 cube in the 1-5 $\mu$m spectral range. This led to acceptable results for sources well detected in the continuum image (i.e. C1, C2, C3). For the other objects we used the [OIII] image instead. The derived $R_e$ values are also given in Table \ref{table:kinematic properties}.  

\begin{figure}[h]
    \centering
    \includegraphics[width=0.49\textwidth]{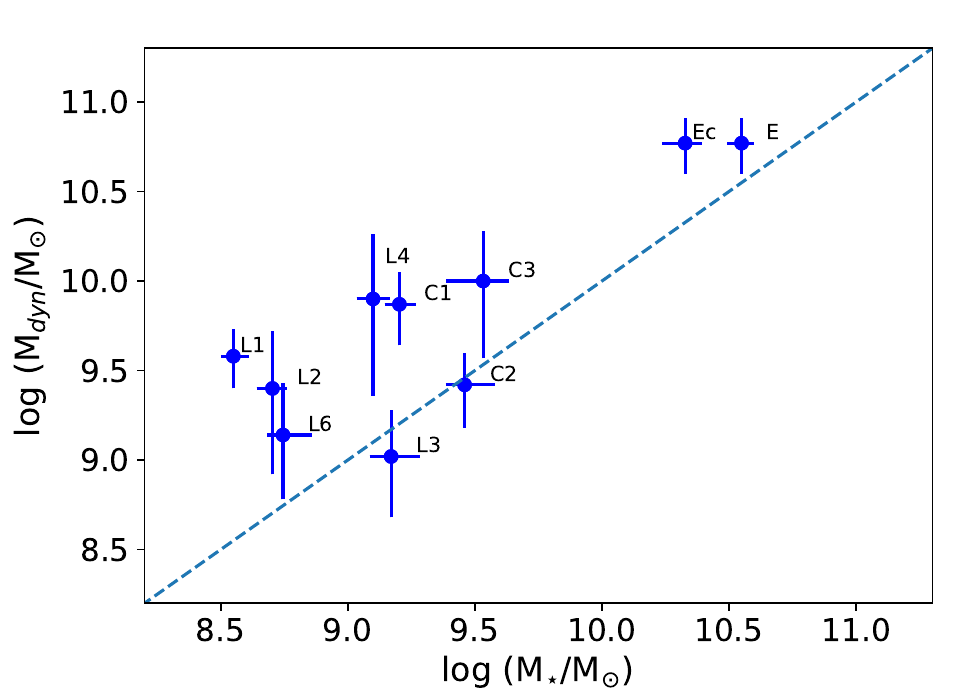}
    \caption{Comparison between the dynamical and the stellar masses for sources in \SPT. For the E galaxy the position for the stellar mass value obtained with CIGALE is indicated with Ec (see sections \ref{subsubsec: kinematics of individual galaxies} and \ref{subsec:Stellar masses}).    
    } 
    \label{fig:Mdyn.vs.Msed}
\end{figure}

The dynamical masses for the newly discovered galaxies are in the range $\sim$ 10$^9$ -- 10$^{10}$ \Msun\ (Table \ref{table:kinematic properties}). Then, although these objects are significantly less massive than the two main galaxies in \SPT, they contain a substantial amount of mass for their redshifts. In Figure \ref{fig:Mdyn.vs.Msed} we compare these dynamical masses with the stellar masses obtained in Sect. \ref{subsec:Stellar masses}. The figure shows that on average $M_{\rm dyn}$ is larger than \Mstar\ by a factor 2-3. This factor may be even higher due to the systematic overestimation of \Mstar\ when MIRI photometry is not included in the SED fitting (\citealp{Papovich2023}; \citealp{Wang2024}). This result suggests that, as expected, these galaxies may have an important fraction of their mass in gas \citep{Tacconi2020ARA&A}. Additionally, or alternatively, they might also have substantial amounts of dark matter, as suggested by recent NIRSpec observations of high-z galaxies \citep{deGraaff2023}. We note that, in Figure \ref{fig:Mdyn.vs.Msed}, we also include the E galaxy, for which  we derive a dynamical mass of 6.0 $\pm$ 2 $\times$10$^{10}$ \Msun.  \cite{Marrone2018} find a total baryonic (i.e. stars, gas, and dust) mass of 7.5 $\pm$ 3.5 $\times$10$^{10}$ \Msun\ and \cite{Alvarez-MarquezSPT} 4-5 $\times$10$^{10}$ \Msun. Therefore, its M$_{\rm dyn}$ determination is consistent with previous results within the estimated uncertainties. 

\begin{table*}
\caption{Kinematic properties and dynamical mass estimates for sources in SPT0311-58}
\label{table:kinematic properties}    
\centering
\begin{tabular}{l c c c c c c  }
\hline
\hline
ID &                 z        &   $\Delta$v       &     $\sigma$      &   R$_e$                     &         M$_{\rm dyn}$             &           Dw    \\
   &                          &  \kms             &      \kms         &   kpc                    &   10$^{9}$\Msun      &           kpc   \\
\hline 
C1&    6.92655 $\pm$   0.00019&  932.  $\pm$    7.&  99.  $\pm$  8. &   0.64  $\pm$   0.19&    7.4$^{+10.}_{-2.7}$ &   9.4  $\pm$    0.3 \\
C2&    6.92419 $\pm$   0.00010&  843.  $\pm$    4.&  57.  $\pm$  5. &   0.69  $\pm$   0.21&    2.6$^{+3.7}_{-1.0}$ &  10.7  $\pm$    0.3 \\
C3&    6.90638 $\pm$   0.00035&  166.  $\pm$   13.& 106.  $\pm$ 15. &   0.75  $\pm$   0.37&    10.$^{+20.}_{-3.7}$ &   3.9  $\pm$    0.5 \\
L1&    6.89941 $\pm$   0.00005&  -98.  $\pm$    2.&  66.  $\pm$  2. &   0.73  $\pm$   0.22&    3.8$^{+4.7}_{-1.4}$&    5.1  $\pm$    0.3 \\
L2&    6.89256 $\pm$   0.00025& -358.  $\pm$    9.&  64.  $\pm$ 12. &   0.51  $\pm$   0.26&    2.5$^{+5.8}_{-0.9}$&    7.6  $\pm$    0.5 \\
L3&    6.88632 $\pm$   0.00011& -595.  $\pm$    4.&  36.  $\pm$  7. &   0.68  $\pm$   0.20&    1.0$^{+2.0}_{-0.4}$&   15.0  $\pm$    0.5 \\
L4&    6.92425 $\pm$   0.00056&  845.  $\pm$   21.&  99.  $\pm$ 24. &   0.70  $\pm$   0.35&    7.9$^{+20.}_{-2.9}$&    8.2  $\pm$    0.5 \\
L5&    6.90940 $\pm$   0.00004&  281.  $\pm$    2.&  36.  $\pm$  3. &          ...        &      ...              &   14.2  $\pm$    0.5 \\
L6&    6.91830 $\pm$   0.00018&  619.  $\pm$    7.&  45.  $\pm$ 10. &   0.58  $\pm$   0.17&    1.4$^{+2.8}_{-0.5}$&    7.3  $\pm$    0.5 \\
L7$^1$&6.92644 $\pm$   0.00054&  928.  $\pm$   20.&  93.  $\pm$ 15. &   0.53  $\pm$   0.31&    5.3$^{+1.8}_{-1.6}$&   11.0  $\pm$    0.5 \\
E &    6.91956 $\pm$   0.00008&  666.  $\pm$    3.& 155.  $\pm$  3. &   2.12  $\pm$   0.63&    59.$^{+70.}_{-21.}$&   10.7  $\pm$    0.3 \\
\hline
\hline
\end{tabular}
\tablefoot{Column 1: Source identification.  
Column 2: Redshift from the R2700 [OIII]-\Hb\ lines. 
Column 3: Relative velocity with respect to z=6.902, as inferred for the W galaxy from the ALMA [CII] line (see Sect. \ref{subsec: Wgalaxy}). 
Column 4: Integrated velocity dispersion derived from the [OIII]-\Hb\ lines in the R2700 integrated spectra, after deconvolving the instrumental profile (Appendix \ref{appendix:spectral_methods}). Column 5: Effective radius estimated as the largest sigma of a 2D Gaussian fit to source emission, after deconvolving quadratically the PSF profile. To derive the properties of the PSF we used the datacube of the standard star (see Sect. \ref{subsec:ancillary}). For sources with significant continuum emission (C1, C2, C3) the sizes were inferred from a panchromatic image generated from the R100 cube in the 1 to 5 $\mu$m spectral range. For the other sources, which have undetected (or very faint) continuum, the [OIII] image was preferred. For the E galaxy the combination of continuum plus [OIII] was used. The Gaussian fits provided unrealistically small errors. Instead, uncertainties are estimated to be 30\%, except for some particularly problematic cases for which 50\% was adopted instead, specifically for C3 (edge of the FoV), L2 (faint), and L4 (badly defined). For L5, which is very close to the edge of the FoV, R$_e$ could not be estimated.  
Column 6:  Dynamical mass of the source adopting the calibration by \cite{Cappellari2006} (see Sect. \ref{subsubsec: kinematics of individual galaxies}). 
Column 7: Distance to spaxel (74,45), which is estimated to be at the centre of gravity of the system. 
$^1$ For L7, the kinematic properties were obtained after the 2D kinematic decomposition of this source and the main component of the E galaxy (Sect. \ref{subsubsection:Decomposition_twocomponentmaps}). }

\end{table*}

\subsection{E galaxy}
\label{subsec:Proto-disc of the E galaxy} 

The kinematic maps for the E galaxy as inferred from the [OIII] lines are presented in the upper panels of Fig. \ref{fig:Kinematics_twocomponentmaps.png}. 
The velocity field shows a significant gradient of $\sim$ 200 \kms along the major axis. The isovelocity lines clearly deviate from what is expected for a regular rotating disc, especially to the northwest where they are strongly twisted. For this spatially extended source, we are able to measure the intrinsic velocity dispersion ($\sigma_0$, moment 2 in Fig. \ref{fig:Kinematics_twocomponentmaps.png}).  Similarly to the velocity field, the $\sigma_0$ map deviates from the expectation for a rotating disc, as it shows local changes at subkpc scales, and higher values at the edges than to the centre. It also shows a general trend along the major axis with, on average, higher values to southeast ($\sigma_0$ $\sim$ 120-145 \kms) than to the northwest ($\sigma_0$ $\sim$ 75-100 \kms). The velocity dispersion from the integrated spectrum (i.e. $\sigma$), which also includes the global velocity gradient, is 155 $\pm$ 3 \kms. The median intrinsic velocity dispersion ($\sigma_0$) over the area covered by Fig. \ref{fig:Kinematics_twocomponentmaps.png} is 113 $\pm$ 19 \kms. From the fits with a 3 $\times$ 3 binning (Fig. \ref{fig:Kinematics}), which cover a larger area but are slightly contaminated by L7, the median $\sigma_0$ = 120 \kms. These values are above the extrapolation of the average $\sigma_0 - z$  evolution found by \cite{Ubler2019}  (i.e. $\sigma_0$ $\sim$ 70-80 \kms at z=7). Interpreting the velocity shear as large-scale rotation and correcting by the inclination suggested by its elliptical morphology, we obtain a rotational support factor, v/$\sigma_0$, of $\sim$ 1,  significantly smaller than in other lower redshift DSFGs (\citealp{Lelli2021}, \citealp{Parlanti2023}, \citealp{Bik2023}, \citealp{Ubler2024}). All these characteristics clearly indicate that the ionised gas disc in the E galaxy in \SPT is not settled yet.

We now compare the kinematic properties of the warm and the cold gas phases, as inferred from NIRSpec and ALMA, respectively. 
The upper-right panel of Fig. \ref{fig:Kinematics} shows that, excluding the region where the source L7 appears in projection, the warm and cold gas components in E have similar velocities, though the latter recedes on average $\sim$ 25 \kms\ faster. 
This behaviour is not homogeneous, and in some spaxels towards the south and east the warm gas has receding velocities of up to $\sim$ +80-100 \kms\ larger.
These differences indicate that the ionised and neutral gas components probed by NIRSpec and ALMA are not cospatial. This may happen when they occupy geometrically distinct regions, and/or as consequence of the dust attenuation. ALMA traces regions over a large range of dust attenuation, while NIRSpec probes the ionised gas that is located in less extincted (likely more external) regions. Hence, relative motions between regions with different dust attenuation could explain the observed velocity offsets. 

As for the velocity dispersion, the cold and warm gas show a significant difference, particularly towards the south and east (see bottom-right panel of Fig. \ref{fig:Kinematics}). Interestingly recent simulations of 4 < z < 9 galaxies by \cite{Kohandel2023} show that the gas velocity dispersion estimates strongly depend on the tracer used. According to this study, while [CII] traces the gaseous disc, H$\alpha$ also includes the contribution from extraplanar ionised gas beyond the disc. This effect is more significant for massive galaxies (i.e. \Mstar > 10$^9$ \Msun), for which $\sigma$ (H$\alpha$) $>$ 2 $\times$ $\sigma$ ([CII]). These predictions are in general agreement with our observations in the E galaxy, though we find significant spatial variations. In the north and in the central regions the velocity dispersion obtained from [CII] and [OIII] are similar, while to the south and east the $\sigma$ ([OIII]) values are larger by factors of up to $\sim$ 2-3. \footnote{We note that we do not find significant differences between $\sigma$ (H$\alpha$) and $\sigma$ ([OIII]) in the spectra of \SPT.}

A similarly relevant difference between the velocity dispersion of warm ionised gas and its cold phase counterpart has been recently reported  by \cite{Parlanti2023} in a z= 4.8 DSFG observed with JWST. However, in that case the difference was found in the central regions of the galaxy, and explained by the feedback effects of an AGN,  a scenario that can hardly be considered for interpreting the observed behaviour in \SPT. In Section \ref{Sec:Discussion}, we discuss further the kinematics in E in connection with other ISM properties. 

\subsection{W galaxy}
\label{subsec: Wgalaxy}
As commented above, the main body of the W galaxy is not detected in a spaxel-by-spaxel basis, and therefore we cannot infer spatially resolved kinematic information with NIRSpec for this galaxy. However, the R2700 integrated spectrum allow us to infer the redshift from the optical emission lines. As the [OIII] lines are faint in this galaxy, we use the \Ha -[NII] complex and the \Hb\ line, and obtain z= 6.907 $\pm$ 0.002. We note that the uncertainty in this case is relatively large because the spectrum is noisy (Fig. \ref{fig:R2700spectra}).

We also derived the redshift from the integrated spectrum of the [CII] line obtained using the same aperture as above.  The [CII] profile in this spectrum departs significantly from a Gaussian. So, rather than derive the redshift fitting the line, we obtain the flux-weighted averaged wavelegth, and obtained z= 6.902 $\pm$ 0.002. This agrees within errors with the  \citealp{Strandet2017} determination (z=6.900 $\pm$ 0.002), who used several CO and [CII] lines and a beam of 2-3\arcsec. We note that because the W galaxy is well detected and resolved with ALMA, and it is also close to the zone of multiple images, its redshift determination may be affected by the lens, which introduces an additional uncertainty.

The differences in the redshifts derived from the optical and the far infrared lines suggest an average shift of $\sim$ +200-300 \kms\ between the warm and the cold gas phases. \cite{Alvarez-MarquezSPT} already reported an offset of Pa$\alpha$ with respect to the [CII] and [OIII] far-IR lines, though in this case significantly larger in amplitude (i.e. +700 \kms). For source C3, at the south of W, the average relative velocity (cold-warm) is $\sim$ +80 \kms. As discussed for the E galaxy, this velocity difference may be explained by the relative motions between the less attenuated regions traced with the optical lines, and the dustier zones probed at the far infrared. The larger attenuation in W (see Sect. \ref{subsec:Attenuation}) could also qualitatively explain why we find a significantly larger velocity offset in this galaxy than in E.

\begin{figure}[h]
    \centering
    \includegraphics[width=0.50\textwidth]{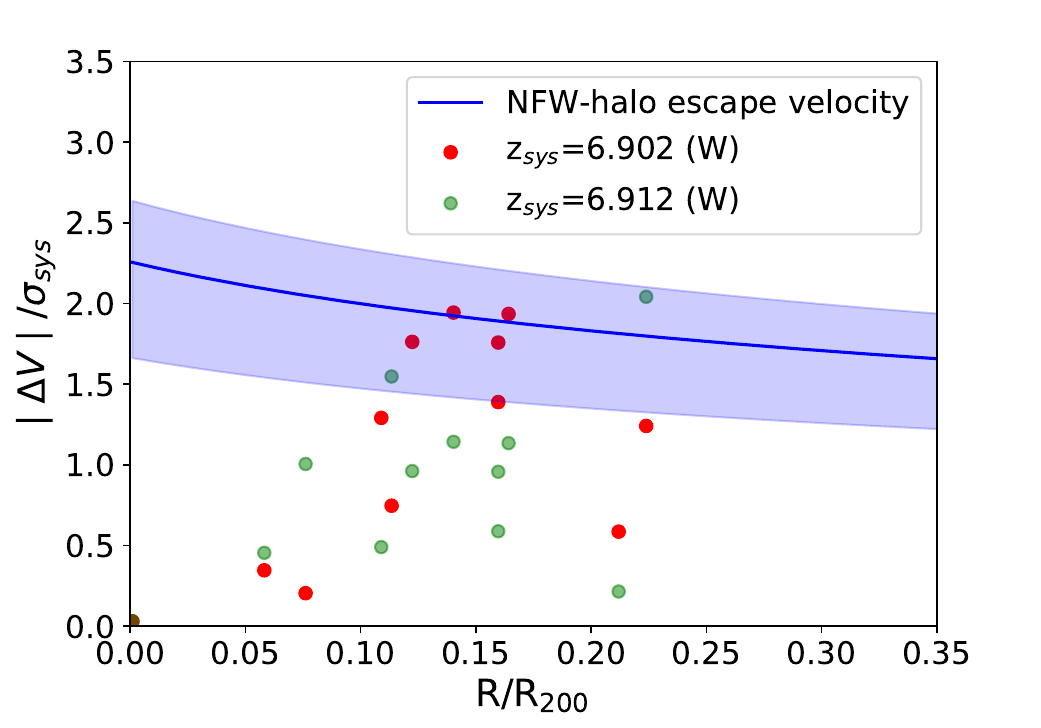}
    \caption{Projected phase space for the galaxies in \SPT.  The x-axis is the projected distance from the cluster centre in units of R$_{200}$. The y-axis shows the absolute value of the relative velocity along the line of sight of each galaxy with respect to the global recessional velocity, normalised by the velocity dispersion of the cluster. The red dots correspond to a global cluster redshift of 
     z$_{sys}$=6.902 as inferred for the W galaxy from the [CII] line, and the green dots are for an {\it ad-hoc} z$_{sys}$= 6.912. The blue line represents the escape velocity normalised by the velocity dispersion of the cluster for a Navarro, Frenk, and White dark matter halo (\citealp{NFW_1996}) of M$_{200}$ = 5 $\times$ 10$^{12}$\Msun, with the blue shadow defining the region for masses between 2 and 8 $\times$ 10$^{12}$\Msun. The point for the W galaxy at (0,0) has been slightly shifted to make it more visible.       
    }
    \label{fig:PPS}
\end{figure}

\section{Discussion}
\label{Sec:Discussion}

\subsection{\SPT: A protocluster core in a massive dark-matter halo}
\label{Sec: Discussion_protocluster}

The possibility that \SPT forms part of a protocluster was proposed by \cite{Wang2021}, who found a significant overdensity of bright submm sources in the field surrounding its two main galaxies (r $\lesssim$ 1.3 Mpc). However, the lack of spectroscopic redshifts prevented Wang and collaborators to confirm this scenario, and classified \SPT as a protocluster candidate. As has been detailed in Section \ref{subsec:Characterization}, the present NIRSpec data have revealed ten new spectroscopically confirmed z $\sim$ 6.9 sources in the IFU FoV (i.e. 17$\times$17 kpc$^2$) in addition to the already known E and W galaxies, which strongly supports the idea that SPT0311-58 is at the core of one of the most distant protoclusters known. Indeed, the number density of sources  ($\phi$ $\sim$ 10$^{4}\ $Mpc$^{-3}$) is orders of magnitude above the expectation for the field ($\phi_{f}$ $\sim$ 5\ Mpc$^{-3}$, for \Mstar > 10$^{7}$ \Msun), according to the observed galaxy stellar mass functions (e.g. \citealp{Conselice2016}). In addition, we find that the sources in \SPT exhibit a relatively large spread in radial velocity ($\Delta$v $\sim$ 1500 \kms, Table \ref{table:kinematic properties}), akin to other high-z protoclusters (z > 4, e.g. \citealp{Chanchaiworawit2019},  \citealp{Hill2020}, \citealp{Calvi2021}, \citealp{Endsley2022}, \citealp{Morishita2023}, \citealp{Hashimoto2023}, \citealp{Scholtz2023_astroph}). This suggests that the mass of the halo must be very large. In the following, we estimate the main properties of the halo, and discuss the dynamical status of the core of this protocluster.

\subsubsection{Main properties of the dark-matter halo}

We estimate the mass of the halo from the stellar mass, which we obtain by combining our results for the E galaxy and the newly discovered sources with the one for the W galaxy determined by \cite{Alvarez-MarquezSPT} (see Sect. \ref{subsec:Stellar masses}). We adopt the mass determinations for E and W that include MIRI photometry. For the newly discovered sources, which have not MIRI detections, we conservatively apply the corrective factor found by  \cite{Papovich2023} and \cite{Wang2024}. The resulting total stellar mass for the system is (1.1 $\pm$ 0.3) $\times$ 10$^{11}$\Msun.  The \Mstar/$M_h$ ratio is not well constrained for the redshift and stellar mass of \SPT, but empirical modeling (UniverseMachine, \citealp{Behroozi2019, Behroozi2020}) and hidrodynamical simulations (MillenniumTNG, \citealp{Kannan2023}) suggest values of $\sim$ 0.015-- 0.03\footnote{Mock galaxy catalogues generated using the so-called UCHUU-UM approach (\citealp{Ishiyama2021}; \citealp{Prada2023}) lead to a \Mstar/$M_h$ ratio range in agreement with the one adopted here (E. Pérez, F. Prada, priv. comm.).}, which leads to $M_h \sim $ (5$\pm$3) $\times$ 10$^{12}$ \Msun. 
  
This result agrees well with the findings of \cite{Marrone2018}, who estimate $M_h$ in \SPT from its gas content (\citealp{Strandet2017}). Specifically, they found that $M_h$ could vary from a conservative lower limit of 1.7 $\times$ 10$^{12}$ \Msun\  to 7.4 $\times$ 10$^{12}$ \Msun, depending on the approach adopted to determine the mass of the gas. However, we note that recently \cite{jarugula_molecular_2021}, using nonlocal thermodynamic equilibrium radiative transfer models, have found that the mass of gas in \SPT is larger than previously estimated. This result suggests that previous gas-based M$_{h}$ determinations should be corrected by factors of 1.5-2.  

Having an estimate for the halo mass, we can now derive the expected size, R$_{200}$ \footnote{R$_{200}$ is defined as the radius within which the mean enclosed mass density is 200 times the critical density of the Universe at the corresponding redshift.}, for a virialised dark matter (DM) halo at z=6.9. By definition  (e.g. \citealp{NFW_1996}):
\begin{equation}\label{eq:m_200}
    M_{200} \equiv \frac{4}{3}\pi  R_{200}^3 200 \rho_c \ ,   
\end{equation}
where M$_{200}$ $\equiv$ M$_{h}$, and $\rho_c$ is the critical density that varies as
\begin{equation}\label{eq:rho} 
   \rho_c = \frac{3 H^2(z)}{8\pi G} \ ,
\end{equation}
with H(z) and G being the Hubble parameter and the gravitational constant, respectively. From these equations, we derive a size of R$_{200}$ = 67$_{-15}^{+9}$ kpc for a halo mass of (5$\pm$3) $\times$ 10$^{12}$ \Msun. 

The centre of mass of this protocluster should be at the W galaxy. This is based on previous works (\citealt{Marrone2018}; \citealt{Alvarez-MarquezSPT}) that have shown that the baryonic mass of this galaxy is about one order magnitude larger than its companion, the E galaxy. The newly discovered galaxies have dynamical masses significantly lower than E (Sect. \ref{subsec:kinematicsanddynamics}), so they should not affect significantly the determination of the centre of gravity of the system (CoG$_{sys}$). We also adopt the redshift of the W galaxy as the mean redshift for the system. As discussed in Sect. \ref{subsec: Wgalaxy}, the redshift for W is subject to  relatively large uncertainties compared with other sources, with significant differences among the values inferred from the optical emission lines (NIRSpec), Pa$\alpha$ (MIRI), and the molecular lines (ALMA), being also possibly affected by the effects of the lens. We adopt the determination based on the [CII] line (i.e. z=6.902 $\pm$ 0.002), because it is not affected by attenuation, and traces a significantly larger amount of mass (see Sect. \ref{subsec: Wgalaxy}).  Therefore, taking the W galaxy as the centre of the protocluster, our NIRSpec observations probe projected cluster-centric distances of up to 15 kpc (see Table \ref{table:kinematic properties}), which correspond to about 0.2 $\times$ R$_{200}$.

\subsubsection{Probing the dynamical state of the protocluster core}

Given that we are observing such an inner region of a protocluster, one might inquire whether it has already undergone (partial) virialisation.  In an attempt to answer this question, we analysed whether the radial velocities of the 12 galaxies may be drawn from a normal distribution, as expected for a system in equilibrium. With this aim, we performed a Kolmogorov-Smirnov (K-S) test that led to a p-value of 0.04, and therefore we conclude that the observed region is not virialised yet. However, we point out that the result of the K-S test is very dependent on the origin of the relative velocities. If we allow a global offset to the radial velocities of 380 \kms, the p-value reaches 0.75, suggesting a normal distribution. This change would imply that the W galaxy is not at the centre of the protocluster, or that its redshift is larger by $\Delta$z $\sim$ 0.01, i.e.  z = 6.912. We note that this redshift falls between the one inferred from the optical lines (z=6.907) and from Pa$\alpha$ (z = 6.918) (Sect. \ref{subsec: Wgalaxy}). Still, for the reasons mentioned above, we argue that most likely the CoG$_{sys}$ is at W and z$_{sys}$ = 6.902, and therefore the system does not seem to be at equilibrium. Nevertheless, in the remainder of the section we also evaluate how the results are affected if an ad hoc z = 6.912 is used as reference.

Using the present kinematic information, we can also assess if the system is at equilibrium by comparing the velocity dispersion observed in \SPT with the one expected for its mass in a virialised cluster. We derive the velocity dispersion in our data by bootstrap, and obtain a mean estimate of $\sigma_{sys}$ = 479 \kms, and a range from 406 to 551 \kms with a 68 percent confidence\footnote{The bootstrap estimate is preferred over the standard deviation (507 \kms), as it is more robust for small samples (\citealp{Beers1990}).}. According to the numerical simulations of virialised clusters by \cite{Evrard2008}, the predicted velocity dispersion at equilibrium for a cluster with a mass of (5 $\pm$ 3) $\times$ 10$^{12}$ \Msun\ is 372$_{-98}^{+65}$ \kms. Since we measure a larger face value of the velocity dispersion, this comparison suggests that the core in \SPT is not virialised yet. However, taking into account the associated uncertainties, the data are also compatible with being (partially) virialised. Alternatively, the \cite{Evrard2008} relation and the observed velocity dispersion can be used to predict the mass for a virialised protocluster. The derived value (1.05$_{-0.41}^{+0.55}$$\times$10$^{13}$\Msun)  is larger than our halo mass estimate, which again suggests that the system has not reached equilibrium. However, as mentioned above, the uncertainties are large and this conclusion lacks robustness.

In order to further analyse the dynamical status of the observed region, in Fig. \ref{fig:PPS} we locate the galaxies in the projected phase space (PPS) diagram (e.g. \citealp{mahajan+2011,Jaffe2015J}). The x-axis of the plot is the projected distance from the cluster centre normalised to R$_{200}$, while the y-axis indicates the (absolute) velocity of each galaxy with respect to the global recessional velocity (v) normalised by the velocity dispersion of the cluster, that is, $\vert$$\Delta$v$\vert$/$\sigma$$_{sys}$. In the plot we also include with a blue line the (projected) escape velocity from a Navarro Frenk and White (NFW, \citealp{NFW_1996}) dark-matter halo of M$_{200}$ = 5 $\times$ 10$^{12}$ \Msun, normalised by $\sigma$$_{sys}$.
The blue shadow region refers to escape velocities for halo masses from 2 to 8 $\times$ 10$^{12}$ \Msun. For our primary reference (i.e. W at the CoG$_{sys}$, z$_{sys}$=6.902, red dots), most of the galaxies are below the escape velocity limit, and therefore they are bound gravitationally to the halo. However, taking into account the uncertainties and the fact that in the figure we present projected velocities corrected with an average factor, a few galaxies could also be consistent with being unbound. When the {\it ad-hoc} redshift (i.e. $z$ = 6.912, see above) is adopted (i.e. green points), one galaxy (L3) appears unbound, though the rest are located deeper into the potential well, showing on average lower $\vert$$\Delta$v$\vert$/$\sigma$$_{sys}$ values. 

In summary, our data suggest that the observed region in \SPT is likely not totally virialised, though it may be not far from reaching equilibrium.  While a few galaxies could be fly-bys, the majority appear to be gravitationally bound to the halo, for which we estimate a DM mass of $\sim$ 5 $\times$ 10$^{12}$ \Msun. These results reinforce the findings by \cite{Marrone2018} who showed that the halo mass for \SPT is very large according the theoretical expectations at z$\sim$ 6.90. 

\begin{figure*}[h]
    \centering
    \includegraphics[width=0.85\textwidth]{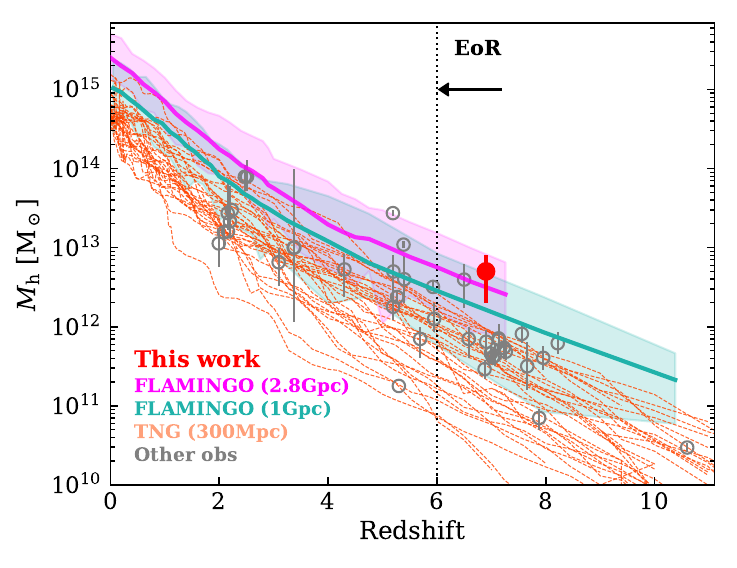}
    \caption{Halo mass of SPT0311-58 together with a compilation of z $>$ 2 protoclusters as a function of redshift (see text). The orange lines correspond to the halo accretion history of the 25 most massive clusters at z=0 in the IllustrisTNG simulation discussed by \cite{Lim2021}. The FLAMINGO simulations are taken from \citet{Lim2024}, and represent the median and the full range among the 100 highest-SFR haloes identified in each snapshot of its 2.8\,Gpc and 1\,Gpc box runs.
    The figure shows the extreme mass of SPT0311-58, being the most massive protocluster discovered so far at EoR. It also shows the difficulty of TNG simulations to predict a rare system like \SPT, as a consequence of their limited simulated volume (i.e. box size of 300\,Mpc). The position of \SPT can be reproduced by FLAMINGO simulations with larger box sizes (see text).}
    \label{fig:fig_mass_comp}
\end{figure*}


\subsubsection{The \SPT protocluster in context: comparison with cosmological simulations and other observations}

The previous findings suggesting an extreme halo mass in \SPT are strengthened further when comparing it with the most massive protoclusters in the IllustrisTNG (e.g. \citealp{Springel2018}) and FLAMINGO \citep{Schaye2023, Kugel2023} cosmological simulations, as well as with previously observed protoclusters.

Figure ~\ref{fig:fig_mass_comp} presents the position of \SPT in the $M_{halo}$-z plane together with the predictions from IllustrisTNG obtained by \citet{Lim2021} by selecting the 25 most massive clusters in the $z\,{=}\,0$ snapshot, and tracking them using the merger tree to find their progenitors at higher redshifts. For FLAMINGO, on the other hand, the figure includes the median and the full range among the 100 highest-SFR haloes identified in each snapshot of its 2.8\,Gpc and 1\,Gpc box runs taken from \citet{Lim2024}. The figure clearly indicates that only the most massive protoclusters in FLAMINGO simulations are able to reproduce the \SPT position. Part of the reason why the protoclusters from IllustrisTNG have significantly lower mass is because of the smaller box size of 300\,Mpc of these simulations, which makes difficult to capture a rare object like \SPT. According to these simulations, protoclusters with masses of ${\gtrsim}\,10^{12}\,{\rm M}_\odot$ at $z\,{\simeq}\,7$ could transform into $10^{15}\,{\rm M}_\odot$ systems at $z\,{=}\,0$, though this may not be necessarily the case (e.g. \citealp{Remus2023}). These results suggest that \SPT, for which independent approaches find a dark matter halo mass higher than $10^{12}\,{\rm M}_\odot$, could be the progenitor of a Coma-like cluster. 

Fig.~\ref{fig:fig_mass_comp} also includes a set of observationally identified protoclusters at $z\,{\gtrsim}\,2$ from the literature, which consists of GOODS-N $z\,{=}\,1.99$ protocluster \citep{Blain2004, Chapman2009}, MRC1138$-$262 \citep{Dannerbauer2014}, SSA22 \citep{Umehata2015}, AzTEC-3 \citep{Capak2011}, SPT2349$-$56 \citep{Miller2018, Hill2020}, ZFIRE \citep{Hung2016}, PHz G237.01$+$42.50 \citep{Polletta2021}, CC2.2 \citep{Darvish2020}, PCL1002 \citep{Casey2015}, CLJ1001 \citep{WangT2018}, ), a LAE overdensity at z=6.5 \citep{Chanchaiworawit2019}, a protocluster in COSMOS \citep{Brinch2024}, MAGAZ3NE J095924$+$022537 and MAGAZ3NE J100028$+$023349 (both from \citealt{McConachie2022}), HDF850.1 \citep{Calvi2021}, z57OD and z66OD \citep{Harikane2019}, an overdensity at $z\,{=}\,7.66$ \citep{Laporte2022}, A2744-z7p9OD \citep{Morishita2023}, the seventeen protocluster candidates identified with JWST by \citet{Helton2023a, Helton2023b}, and GNz11 (\citealp{Scholtz2023_astroph}). This illustrates the distinct mass of SPT0311$-$58 among the protoclusters discovered at high redshifts, being currently the most massive known at the EoR. 

The present study, although admittedly subject to large uncertainties, represents one of the first attempts to characterise spectroscopically a protocluster at the epoch of reionisation, and it shows the power of JWST for such studies. Further observations expanding the observed area will enable us to better characterise the field, and constrain the dynamical stage of this extreme system.

\subsection{The turbulent early phases of galaxy assembly: accretion, mergers, and gravitational interactions}
\label{subsubsect:dicussion_turbulent_disc}

In this section, we discuss the ongoing physical processes and the current evolutionary stage of the E galaxy, based on the detailed spatially resolved properties provided by the NIRSpec IFU  presented in previous sections.

\subsubsection{Metallicity gradient and evidence of gas accretion in the E galaxy}

As shown in Fig. \ref{fig:Metal_map.png},  
the E galaxy shows a significant metallicity gradient of $\sim$ 0.7 dex, with the lowest values (i.e. 12-log(O/H) $\sim$ 7.9) found in the northwest region. This is close to the zone where the H$\alpha$ and [CII] emission extends towards the position of the C1 galaxy (see grey isocontours in Fig.\ref{fig:Metal_map.png}, and  Fig. \ref{fig:R100imaging}(c,f)), indicating the presence of significant amounts of warm and cold gas. In this region, the stellar component, as traced by the continuum maps (panels a and d in Fig. \ref{fig:R100imaging}), is not prominent. This indicates that the outskirts of E in the northwest direction should be dominated by gas. Moreover, the velocity field in the northwest part of E shows significantly twisted isolines that cannot be explained as part of a global regular rotation (Sec. \ref{subsec:kinematicsanddynamics}). All these features can be naturally explained as the result of the gas accretion of low-metallicity gas from the CGM into the main E structure. 

In lower redshift galaxies (i.e. z$\sim$ 3) metallicity gradients have also been found, and explained by the accretion of primordial gas (\citealp{Cresci2010_metalgradientsinflows}) as predicted by theoretical models (\citealp{Dekel_2009Natur.457..451D}).  The gas accretion in \SPT could be triggered by the interaction of E with the C1 galaxy, which is only $\sim$ 100 \kms\ and 2 kpc apart in projection. According to the velocity field of E and the radial velocity of C1, this possible encounter would result in a prograde interaction, which is an efficient process at driving gas inflows (\citealp{BarnesHernquist1992}, \citealp{Cox2009}, \citealp{Lambas2012}).  Alternatively, E and C1 may not be interacting (i.e. simply appear close in projection), and the accretion of low-metallicity gas into E could be the result of smooth streams of gas from the IGM (\citealp{SanchezAlmeida2014}). This possibility is consistent with cosmological simulations that predict that streams of inflowing gas from the cosmic web are the dominant mechanism for the formation of galaxies in massive halos (e.g. \citealp{Dekel_2009Natur.457..451D}, \citealp{Ceverino2016_inflow_metaldrops}). As we discuss in the previous section, \SPT resides in a massive halo and therefore the presence of streams feeding its central regions with gas from the IGM is naturally expected when it is actively building its stellar component at rates $>$ 100s \sfr. 

The observed metallicity gradient could, in principle, also be explained by the merging of two (or more) galaxies with a significant difference in metallicity. The continuum maps (Fig. \ref{fig:F125W}, Fig. \ref{fig:R100imaging}(a)) show two distinct structures in the E galaxy, i.e. ES and EN, which could be the result of two separate precoalesce merging galaxies. The MIRI image (Fig. \ref{fig:R100imaging}d) also shows a structure to the north-west that is not seen in NIRSpec likely due to obscuration. However, the kinematic properties in the main structure of E do not show evidence for two kinematically distinct objects, such as peaks in the velocity dispersion map, or rotation associated with their nuclear regions. The [OIII] (Fig. \ref{fig:Kinematics_twocomponentmaps.png} upper panels) and [CII] (Fig. \ref{fig:Kinematics}) velocity fields also show a smooth behaviour at large scales. The most significant distortion in the velocity map is found at the northwest edge of the main structure (i.e. not in an inner region), but the velocity dispersion there is relatively low, which does not suggest gas motions associated with strong interactions. Therefore, while we cannot rule out completely the merging scenario, the ionised gas, metallicity, and kinematic maps presented here favour the gas accretion scenario in the E galaxy.  

A third alternative to explain metallicity gradients is the presence of outflows (e.g. \citealp{RodriguezdelPino2023arXiv}). However, we do not see any clear evidence for outflows in our spectra. Moreover, if an outflow were present (or had been active in recent times), its most likely orientation would be along the minor axis, which is perpendicular to the direction of the metallicity gradient. Therefore, the outflow scenario can hardly explain the case of \SPT.  

\subsubsection{Kinematic evidence for an unsettled disc structure at redshift $\sim$ 7}

We now discuss the properties of L7, and the possibility that it will merge with E. The kinematic maps for these two galaxies are shown in Fig. \ref{fig:Kinematics_twocomponentmaps.png}. 
One clear feature that can be inferred from the figure is that L7 is reshifted by $\sim$ 250 \kms with respect to E. This fact has two implications. First, this significant kinematic decoupling rules out that L7 has been formed by disc fragmentation in contrast to the clumps identified by \citet{Spilker2022} at submm wavelengths. Second, the relative velocity between L7 and E allows us to constrain the geometry. In fact, taking into account the dusty nature of E, L7 has necessarily to be located between us and the main structure of E, as otherwise it could not be observed through the dusty disc. Therefore its receding velocity suggests that it is  falling into E. Although we cannot rule out that L7 has a large velocity component perpendicular to the line of sight, its already large projected velocity with respect to W (i.e. +928 \kms) makes this possibility unlikely (i.e. otherwise, its actual velocity with respect to W would be extremely high). So, rather than a fortuitous alignment of a fly-by source, the data suggest that we are witnessing the premerger phase between L7-E. The mass ratio of both galaxies is $\lesssim$ 1:8, and thus it should be classified as a minor merger event, the likes of which are expected to occur frequently according to theoretical models (e.g. \citealp{Ceverino2016_inflow_metaldrops}).      

In Sect. \ref{subsec:kinematicsanddynamics} we show that, in addition to the irregular velocity field, the velocity dispersion map of the E galaxy exhibits a nonaxisymmetric structure with variations at subkpc scales and a global trend along the northwest - southeast direction (Fig. \ref{fig:Kinematics_twocomponentmaps.png}). The velocity dispersion is high everywhere, with values ranging from $\sim$ 75 up to 150 \kms, indicating high turbulence.
The velocity dispersion map does not show the expected increase due to beam-smeared velocity gradients in the central regions, as observed in discs formed
at later cosmic times (e.g. \citealp{Ubler2019}). The sparse isovelocity lines shown in Fig. \ref{fig:Kinematics_twocomponentmaps.png} clearly indicate that the situation in this galaxy is rather different. 

Then, the present 2D kinematic maps (i.e. both the velocity field and the velocity dispersion map) describe an ionised gas structure that departs from what is expected for a settled rotating disc. The kinematics of the cold gas component obtained with high-resolution [CII]$\lambda$158 $\mu$m ALMA observations (\citealp{Spilker2022}), also reveal high velocity dispersions associated with the clumps, indicating large turbulence. The mean radial velocities of the clumps also follow a global velocity gradient, and thus they could be the result of disc fragmentation. Hence, the main kinematic properties of the E galaxy, as inferred from the ionised and neutral gas phases,  differ significantly from the settled and regular behaviour observed in some DSFGs at lower redshifts (i.e. z $\sim$ 4-5, \citealp{Fraternali2021}; \citealp{Rizzo2021}; \citealp{Lelli2021}; \citealp{Parlanti2023}; \citealp{Ubler2024}).
In fact, the E galaxy is actively assembling its stellar component undergoing a highly turbulent phase marked by the presence of gravitational instabilities, gas accretion, and merger events. All of these processes also account for the irregularities and subkiloparsec variations in the interstellar medium (ISM) conditions, as evidenced by the ionisation and metallicity maps.

Still, despite these tumultuous conditions, the velocity field in E shows a clear large-scale velocity gradient of about 200 \kms. If interpreted as global rotation, the E galaxy will have several complete revolutions ($\sim$ 4 at r = 3 kpc) during the next $\sim$ 0.5 Gyr (i.e. by z $\sim$ 4.5). Together with other mixing mechanisms, this will contribute to redistributing the gas and erase the physical and kinematic inhomogeneities produced by present and recent events. Therefore, it is conceivable that this process will make the properties of the E galaxy in \SPT to evolve into those of more settled DSFGs currently observed at z $\sim$ 4-5 (e.g. \citealp{Rizzo2021}, \citealp{Parlanti2023}, \citealp{Ubler2024}), unless additional accretion and/or mergers with nearby objects continue to generate inhomogeneities in the ISM and disturb its kinematics.        

\section{Summary and conclusions}
\label{section:summary}

We present JWST/NIRSpec IFU low-resolution (R=100) and high-resolution  (R=2700) spectroscopy of SPT0311-58 at z $\sim$ 6.9, the most massive and distant dusty star-forming system known, and a protocluster candidate. The main conclusions of this paper can be summarised as follows:

(i) Our observations reveal ten new sources at a redshift of  $\sim$ 6.9 (and three additional foreground galaxies at z $\sim$ 1-3) in the field around the two  main, previously identified massive galaxies of the system, referred to as the E and W galaxies. The newly discovered z $\sim$ 6.9 galaxies are characterised by dynamical masses of $\sim$ 10$^9$-10$^{10}$ \Msun, and a wide range in relative velocity ($\Delta$v $\approx$ 1500 \kms). The high density of sources ($\phi$ $\sim$ 10$^{4}\ $Mpc$^{-3}$) and their spread in radial velocity confirm that \SPT is at the core of an extremely massive protocluster, for which we estimate a mass of $\sim$ (5 $\pm$ 3) $\times$ 10$^{12}$ \Msun\ from its stellar content. \SPT\ is therefore the most massive protocluster discovered so far at the EoR. A basic analysis indicates that the innermost regions of the protocluster (r < 0.2 R$_{200}$) have likely not yet reached virial equilibrium. We also located the galaxies in the projected phase space (PPS) diagram, and find that most are gravitationally bound to the halo. According to cosmological simulations  for a protocluster with its mass, it could evolve into a Coma-like system at z=0, though this may not be necessarily the case.

(ii) The ISM conditions of the sources in \SPT are very diverse. While some of these z $\sim$ 6.9 objects have similar ionisation properties to those of the field galaxies at these redshifts recently observed with JWST, others have conditions mostly found in more evolved objects at lower redshift. In the N2-BPT diagnostic diagram, most sources occupy the region of composite SF/AGN ionisation at low-z, but we do not find clear evidence for AGN. None are in the local LI(N)ER region, but shocks are likely to be present in a turbulent ISM, which is expected to experience disc fragmentation, interactions,  and minor mergers. Similarly, the metallicity varies by more than 0.8 dex across the galaxies in the FoV. 
The regions with the highest metallicity are associated with the two main massive galaxies, with nearly solar values of $\sim$ 8.5-8.6 in the 12+log(O/H) scale. Three external sources show the lowest metallicities with values of about 7.9-8.1.

(iii) Using H$\alpha,$ we  characterised the unobscured star formation (SFR(H$\alpha$)) in the system, and identified three distinct SF regimes. First, the most massive W galaxy (M$_{bar}$ $\sim$ 3 $\times$ 10$^{11}$ \Msun) has a very modest  SFR(H$\alpha$) ($\sim$ 8 \sfr) compared with the enormous obscured SF detected with ALMA (i.e. SFR(FIR) $\sim$ 3000 \sfr). Most of the SF in this galaxy is totally blocked at optical and near-IR wavelengths, and cannot be recovered with the dust-attenuation correction. Second, the less massive, though still very massive (M$_{bar}$ $\sim$ 4 $\times$ 10$^{10}$ \Msun) E galaxy, has significant SFR(H$\alpha$) $\sim$ 80 \sfr,  with a dominant SFR (FIR) of $\sim$ 500 \sfr. The dust-corrected SFR (H$\alpha$) recovers most of the total SFR of the system. Finally, the newly discovered sources have SFR(H$\alpha$) $\sim$ 1-6 \sfr, and no obscured SF is detected with ALMA. Their contribution to the total SF of the system is minor ($\sim$ few percent).

(iv) The 2D physical, chemical, and kinematic information of the E galaxy provided by the present data allows us to characterise the current early phases of its stellar mass assembly. The velocity field and the velocity dispersion maps show asymmetric and irregular patterns indicating clear departures from rotation and high turbulence (intrinsic velocity dispersion, $\sigma_0, \sim $ 115 \kms ). We find large-scale motions between the ionised gas component probed with NIRSpec and the cold gas phase counterpart traced by ALMA. 
This galaxy shows a clear metallicity gradient of $\sim$ 0.1 dex/kpc that is naturally explained by accretion of pristine gas from the IGM. We also identify a possible (precollision) minor merger event.  All these results suggest a scenario in which the E galaxy is undergoing a tumultuous phase while actively assembling its stellar component. However, the large global velocity gradient observed suggests that it could evolve into a regular settled rotating DSFG, similar to those observed at later cosmic times (i.e. z $\sim$ 4-5).  
 
The present study highlights the role of \SPT as an ideal astrophysical laboratory for studying galaxy evolution in a very massive dark-matter halo at the epoch of reionisation. It also demonstrates the enormous potential of the JWST/NIRSpec IFU for detailed spatially resolved analyses at these very early cosmic times.

\begin{acknowledgements}


We thank an anonymous referee for constructive comments. We also thank Daniel Ceverino, Evencio Mediavilla, Charles Proffitt, Francisco Prada, Enrique Pérez, Irene Shivaei, Rhea-Silvia Remus, Jorge Jiménez-Vicente, and Jose Alberto Rubiño for uselful discussions, and valuable input. 

This work is based on observations made with the NASA/ESA/CSA \textit{James Webb} Space Telescope. The data were obtained from the Mikulski Archive for Space Telescopes at the Space Telescope Science Institute, which is operated by the Association of Universities for Research in Astronomy, Inc., under NASA contract NAS 5-03127 for JWST. 

SA, MP, and BRP  acknowledge support from the research project PID2021-127718NB-I00 of the Spanish Ministry of Science and Innovation/State Agency of Research (MICIN/AEI//10.13039/501100011033). IL acknowledges support from the Spanish Ministry of Science and Innovation (MCIN) by means of the Recovery and Resilience Facility, and the Agencia Estatal de Investigación (AEI) under the projects with references BDC20221289 and PID2019-105423GA-I00. PGP-G acknowledges support from grants PGC2018-093499-B-I00 and PID2022-139567NB-I00 funded by Spanish Ministerio de Ciencia e Innovación MCIN/AEI/10.13039/501100011033, FEDER, UE. RM, FSE, JS, JW, and SL acknowledge support by the Science and Technology Facilities Council (STFC), by the ERC through Advanced Grant 695671 ``QUENCH'', and by the
UKRI Frontier Research grant RISEandFALL. RM also acknowledges funding from a research professorship from the Royal Society.
GC acknowledges the support of the INAF Large Grant 2022 ``The metal circle: a new sharp view of the baryon cycle up to Cosmic Dawn with the latest generation IFU facilities''. 
H{\"U} gratefully acknowledges support by the Isaac Newton Trust and by the Kavli Foundation through a Newton-Kavli Junior Fellowship.
AJB, GCJ, and JCh  acknowledge funding from the ``FirstGalaxies'' Advanced Grant from the European Research Council (ERC) under the European Union’s Horizon 2020 research and innovation program   (Grant agreement No. 789056). SC, EP, and GV acknowledge support from the European Union (ERC, WINGS,101040227). The Cosmic Dawn Center (DAWN) is funded by the Danish National Research Foundation under grant DNRF140.
\\  

This paper makes use of the following ALMA data: 
ADS/JAO.ALMA$\#$2016.1.01293.S and 
ADS/JAO.ALMA$\#$2017.1.01423.S.
ALMA is a partnership of ESO (representing its member states), NSF (USA) and NINS (Japan), together with NRC (Canada), MOST and ASIAA (Taiwan), and KASI (Republic of Korea), in cooperation with the Republic of Chile. The Joint ALMA Observatory is operated by ESO, AUI/NRAO and NAOJ.
This research has made use of "Aladin sky atlas" developed at CDS, Strasbourg Observatory, France \citep{Bonnarel2000}

This research has made use of NASA's Astrophysics Data System, QFitsView, and SAOImageDS9, developed by Smithsonian Astrophysical Observatory.


This research made use of Astropy,  a community-developed core Python package for Astronomy \citep{astropy},  {\tt Matplotlib} \citep{Hunter2007}, {\tt NumPy} \citep{VanDerWalt2011}. 
This research has made use of "Aladin sky atlas" developed at CDS, Strasbourg Observatory, France \citep{Bonnarel2000}.

\end{acknowledgements}

\bibliographystyle{aa}
\bibliography{main}

\begin{appendix}

\section{Spectral characterisation: Methods}
\label{appendix:spectral_methods}

\subsection{Definition of extraction apertures for regional integrated spectra}
\label{subsec:extractionapertures}

\begin{figure*}[h]
    \centering
    \includegraphics[width=18.cm,trim= 0 0 0 0,clip]
    {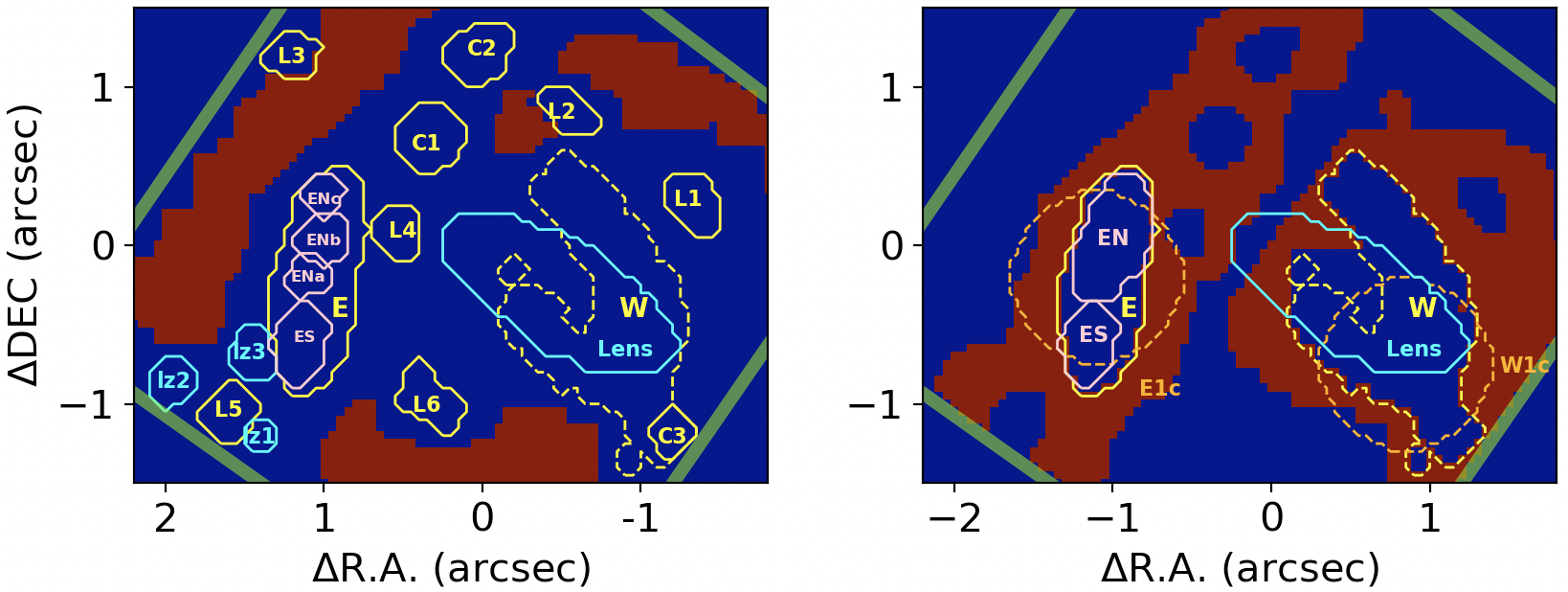}
    \caption{Selected apertures. Left panel:  Yellow lines indicate the apertures for the z $\sim$ 6.9 sources, while with cyan is used for the low-z objects, including the lens galaxy. For the W galaxy we use a dashed line to indicate that this aperture is not defined on the basis of the NIRSpec data, but from the 160 $\mu$m continuum ALMA image. Apertures for the subregions within the E galaxy are indicated with pink lines. The red region in this panel marks the region used to obtain the background spectrum. Right panel: Additional apertures, including the ones with 1 arcsec in diameter (i.e. E1c and W1c, in orange lines) used by \cite{Alvarez-MarquezSPT}. In this panel the red regions represent the spaxels used to probe the diffuse external emission outside the E and W galaxies (i.e. the E$_{ext}$ and W$_{ext}$ apertures). 
    }
    \label{fig:Apertures}
\end{figure*}


We defined a set of extraction apertures, which were used to obtain the integrated spectra for the sources in \SPT (Fig. \ref{fig:Apertures}). The definition of these apertures is, in some cases, difficult because the emission distribution of the lines and the continuum is different, and also because the presence of artefacts force us to exclude some spaxels from the aperture that would have been otherwise included. As a consequence, the absolute fluxes inferred from the spectra should be taken with caution, also considering that the calibration of the instrument is not consolidated yet. However, the velocity determinations, and the flux line ratios should be reliable within the uncertainties imposed by the S/N. In addition, we obtained and applied specific aperture corrections to the individual apertures, which (partially) correct the derived absolute fluxes (see next section). 

Figure \ref{fig:Apertures} shows the apertures used. On the left panel, yellow lines indicate the apertures for the z $\sim$ 6.9 sources. For the W galaxy we use a dashed line to indicate that this aperture is not defined on the basis of the NIRSpec data, but from the 160 $\mu$m continuum ALMA image. Subregions within the E galaxy are indicated with pink lines. For the low-z objects the apertures are displayed with cyan lines (see also Fig. \ref{fig:schematic_apertures.png} for a simplified version of this figure). The red region in this panel represents the region used to obtain the background spectrum. In the right panel, we show additional apertures, including the ones with 1 arcsec in diameter (i.e. E1c and W1c, in orange lines) used to compare with the Pa$\alpha$ measurements obtained with MIRI (\cite{Alvarez-MarquezSPT}). In this panel the red regions represent the spaxels used to probe the external diffuse external emission outside the E and W galaxies (i.e. the E$_{ext}$ and W$_{ext}$ apertures). These extended apertures were only used to probe the H$\alpha$ emission at the redshift of \SPT, so the fact that for the W it includes some emission from the lens galaxy is not critical. Nevertheless, as it can be seen in the figure, we exclude the spaxels with the brightest emission from the lens galaxy to minimise the noise.
 
\subsection{Aperture corrections}
\label{subsubsec:aperture_corrections}

We obtained the aperture corrections making use of the high-resolution G395H/290LP observations of the standard star 1808347 (PID:1128, Obs. 9; PI: N. Luetzgendorf), which were reduced using the same pipeline version and context as for the \SPT data. The aperture masks were shifted such that they were centered with the peak of the star emission. Then we define {\it filters} of about 200 spectral pixels around two wavelengths (3.4 and 5.2 $\mu$m), and the fractions of light contained within the aperture with respect to the total in the FoV were computed. The results are summarised in Table \ref{tab:aperture_corrections}. We note that for E$_{ext}$ and W$_{ext}$, defined for tracing the extended emission, the aperture correction were not obtained. For L7, characterised with a 2D spectral decomposition, the aperture correction is not either computed. 

\begin{table}
\caption{Aperture positions and corrections}   
\centering
\begin{tabular}{l c c c c c c}
\hline
\hline
ID &  X & Y & $\Delta$$\alpha$ ('') & $\Delta$$\delta$ ('') & AC-3.4 & AC-5.2 \\
\hline
E              & 35 & 52 & +1.00 & -0.15 & 0.87 & 0.83 \\
E$_{cir}$$^1$  & 34 & 52 & +1.05 & -0.15 & 0.92 & 0.89 \\
ESb            & 33 & 43 & +1.10 & -0.60 & 0.72 & 0.65 \\
EN             & 36 & 57 & +0.95 & +0.10 & 0.81 & 0.75 \\
ENa            & 34 & 52 & +1.05 & -0.15 & 0.50 & 0.44 \\
ENb            & 36 & 57 & +0.95 & +0.10 & 0.63 & 0.55 \\
ENc            & 36 & 62 & +0.50 & +0.15 & 0.46 & 0.40 \\
W              & 73 & 41 & -0.90 & -0.70 & 0.89 & 0.86 \\
W$_{cir}$$^1$  & 73 & 41 & -0.90 & -0.70 & 0.92 & 0.89 \\
\hline
C1             & 49 & 70 & +0.30 & +0.75 & 0.76 & 0.70 \\
C2             & 55 & 80 & +0.00 & +1.25 & 0.72 & 0.66 \\
C3             & 80 & 32 & -1.25 & -1.15 & 0.51 & 0.45 \\
L1             & 83 & 61 & -1.40 & +0.30 & 0.64 & 0.58 \\
L2             & 67 & 73 & -0.60 & +0.90 & 0.55 & 0.49 \\
L3             & 32 & 80 & +1.15 & +1.25 & 0.60 & 0.54 \\
L4             & 45 & 58 & +0.50 & +0.15 & 0.62 & 0.56 \\
L5             & 24 & 35 & +1.55 & -1.00 & 0.68 & 0.61 \\
L6             & 49 & 36 & +0.30 & -0.95 & 0.59 & 0.56 \\
\hline
lz1            & 28 & 32 & +1.35 & -1.15 & 0.34 & 0.27 \\
lz2            & 17 & 39 & +1.90 & -0.80 & 0.54 & 0.48 \\
lz3            & 27 & 42 & +1.40 & -0.65 & 0.61 & 0.54 \\
\hline
\hline
 
         & 
    \end{tabular}
    \tablefoot{Column 1: Aperture identification, Columns 2 and 3: X and Y coordinates in the cube for the central spaxel of the aperture, Columns 4 and 5: relative sky coordinates with respect to the centre of the FoV (i.e. spaxel [55,55], sky coordinates: $\alpha$ = 03$^h$ 11$^m$ 33$^s$.248 and $\delta$ = -58º 23' 33''.24).  Columns 6 and 7: Aperture corrections at 3.4 $\mu$m  and 5.2 $\mu$m, which correspond approximately to the observed wavelength of H$\gamma$ and H$\alpha$ at z=6.90. $^1$ Circular aperture of 1 arcsec in diameter. 
}
    \label{tab:aperture_corrections}
\end{table}

\subsection{High resolution (R2700) integrated spectra}
\label{subsubsec:R2700_spectra}

\begin{figure*}[h]
    \includegraphics[width=0.95\textwidth,trim= 0 0 0 0,clip]
    {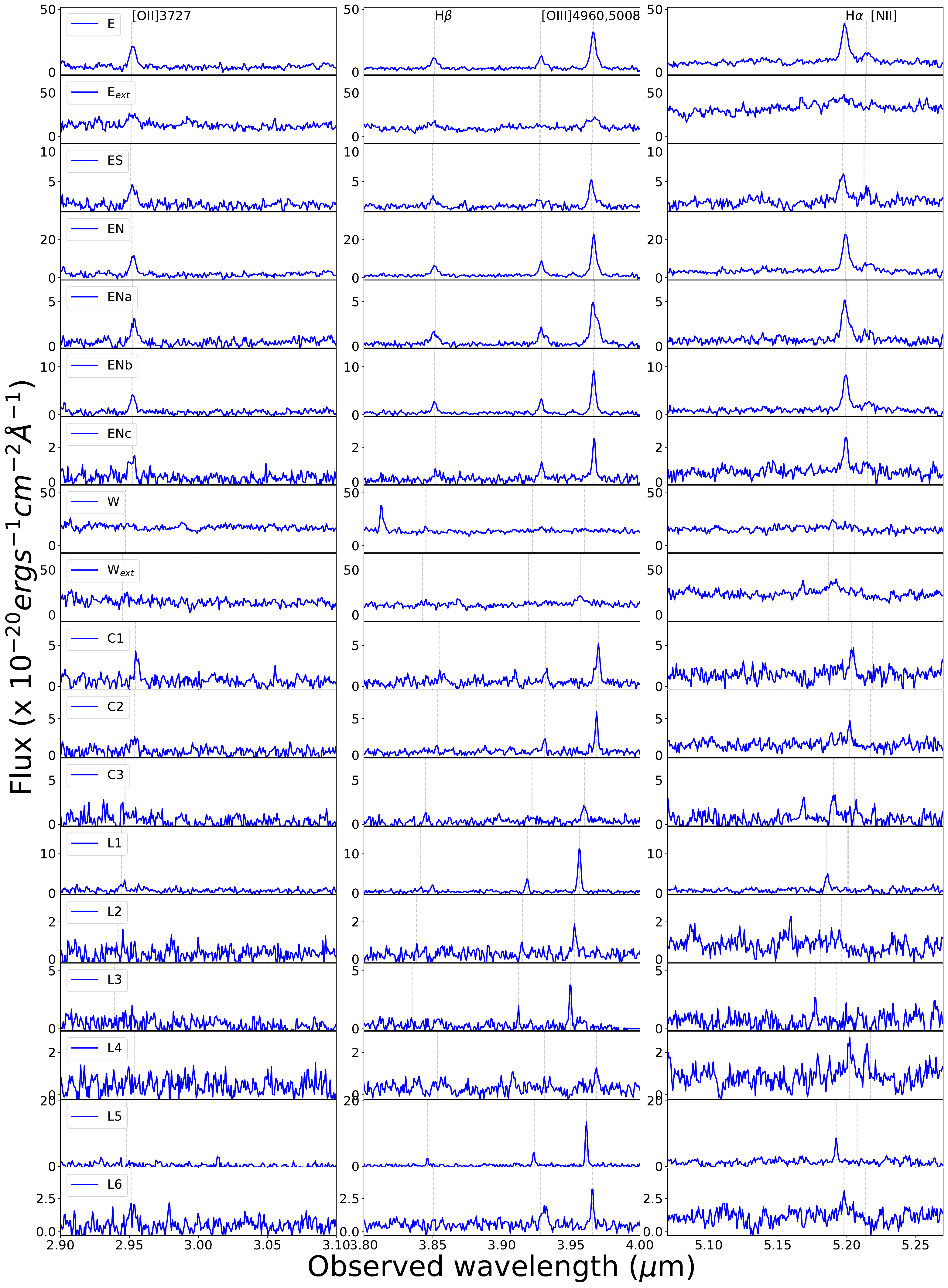}
    \caption{R2700 spectra in the region of the [OII]$\lambda$$\lambda$3726,3728, H$\beta$-[OIII]$\lambda$$\lambda$4959,5007, and H$\alpha$-[NII]$\lambda$6584 for the sources and regions in SPT0311-58. The vertical lines mark the central position of these lines at the redshift of the source.  
    }
    \label{fig:R2700spectra}
\end{figure*}

The integrated R2700 spectra for the sources and regions in \SPT are presented in Fig. \ref{fig:R2700spectra}.  The R100 spectra were presented in Fig. \ref{fig:R100_spectra}.

\subsection {Line fitting}
\label{subsubsec:line fitting}

The emission lines in the R100 and R2700 integrated spectra were fit to single Gaussian profiles linking their relative fluxes and wavelengths according to their atomic parameters, and imposing the same line width for all the lines. The spectral ranges [OII], \Hb-[OIII], and \Ha-[NII] were fitted independently, though for the [OII] and \Ha-[NII], the kinematic parameters were constrained to the one derived for the \Hb-[OIII] complex.   
   
In the  R100 spectra \ha\ and the [NII] lines are not well resolved and the results of the fit may be degenerated, especially at low S/N. The [OII]$\lambda\lambda3726,3728$ doublet is unresolved at R2700 and fit with a single Gaussian, as in this case we were only interested in the total flux (i.e. not in the kinematics). 

For the regions with the highest S/N (i.e. E galaxy), line fitting at spaxel level (or with moderate binning) was performed. The R2700 spectra were used to disentangle two kinematically distinct components following the approach described in \citet{Perna2022Fits} (see Sect. \ref{subsubsection:Decomposition_twocomponentmaps}).

\subsection {Line fluxes}
\label{subsubsec:line fluxes}

Table \ref{table:line_fluxes} gives the line fluxes and their uncertainties obtained from the Gaussian fits of the R100 and R2700 spectra for the [OII]$\lambda\lambda3726,3728$, H$\beta$, [OIII]$\lambda5008$, H$\alpha$, and [NII]$\lambda6584$ lines. 

\begin{figure*}[h]
    \centering
    \includegraphics[width=0.7\textwidth]{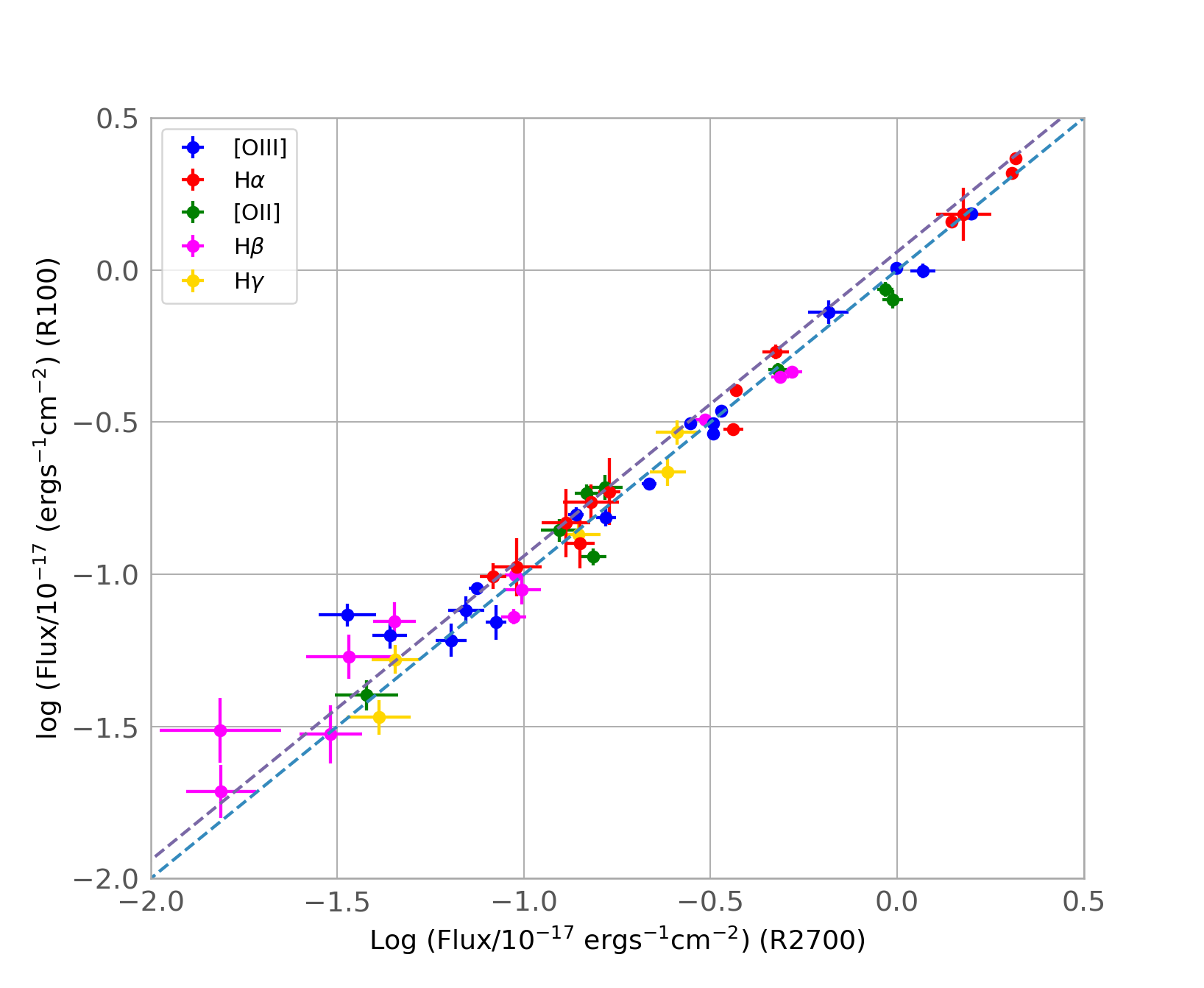}
    \caption{Comparison of the line fluxes obtained from R100 and R2700 integrated spectra. Fluxes from R100 spectra are reduced by 10 percent (see text). The dashed lines correspond to the 1:1 and 1:1.1 ratios. Colours distinguish between lines, and green text labels indicate the aperture/spectrum. 
    } 
    \label{fig:Flux}
\end{figure*}

As the main lines used in this work appear in both the R100 and R2700 datacubes, this give us the opportunity to compare the derived fluxes with the two spectral configurations (see Fig. \ref{fig:Flux}). As it is shown in the Figure, the fluxes span more than two orders of magnitude and indicate in general good agreement. There is however a small systematic trend in the sense that R100 fluxes are about 10 percent higher than R2700 ones. As the R100 data are subject to larger uncertainties in the absolute calibration (Charles Profitt, priv. comm.), we applied this 10 percent correction to the R100 data. After taking into account this factor fluxes typically agree better than 20 percent, except few cases. There is not a configuration (i.e. R100 or R2700) that is generally preferred for obtaining the fluxes. Their spectra have S/N that in many cases are comparable, though they may differ on other data quality aspects. For instance, the H$\alpha$ line at the redshift of \SPT is very close the edge of the spectra, leaving in the case of R100 few pixels for the determination of the continuum redwards the line. Moreover, in R100 spectra H$\alpha$ and the [NII] lines are not resolved, which makes the fits degenerated especially in a low S/N regime. On the other hand, the R2700 datacube is affected in some cases by artefacts that could not be removed by the pipeline nor by a posterior processing. Table \ref{table:line_fluxes} gives further details on the derived fluxes for the lines.

\begin{table*}
\caption{Line fluxes for the sources and regions in the FoV$^1$}
\label{table:line_fluxes}
\begin{tiny}
\begin{tabular}{l c c c c c c c c c c }
\hline \hline

Sp & OII-R100    & OII-R2700   & \Hb-R100    & \Hb-R2700   &  OIII-R100   & OIII-R2700   & \Ha-R100     & \Ha-R2700    & NII-R100    & NII-R2700\\
\hline
E  &8.71$\pm$0.48&8.52$\pm$0.43&4.50$\pm$0.17&4.46$\pm$0.24&15.49$\pm$0.19&14.48$\pm$0.28&21.03$\pm$0.33&18.59$\pm$0.56&5.54$\pm$0.27&4.52$\pm$0.48\\
E1c&8.08$\pm$0.56&8.91$\pm$0.56&4.65$\pm$0.19&4.77$\pm$0.29&15.46$\pm$0.22&14.47$\pm$0.34&22.91$\pm$0.52&18.57$\pm$0.71&5.24$\pm$0.43&4.12$\pm$0.61\\
E$_{ext}$ &             &             &             &             &10.04$\pm$0.56&10.77$\pm$0.83&15.38$\pm$1.51$^2$&13.82$\pm$2.34&         &0.14$\pm$2.64\\
ES&1.95$\pm$0.18&1.51$\pm$0.17&0.70$\pm$0.08&0.70$\pm$0.08&2.00$\pm$0.09 &1.98$\pm$0.09 &3.88$\pm$0.11 &3.09$\pm$0.25 &0.98$\pm$0.09&1.25$\pm$0.23\\
EN &4.74$\pm$0.24&4.39$\pm$0.25&2.81$\pm$0.10&2.42$\pm$0.14&10.26$\pm$0.11&9.14$\pm$0.16 &11.88$\pm$0.20&10.49$\pm$0.32&3.33$\pm$0.17&1.85$\pm$0.27\\
ENa&1.15$\pm$0.07&1.41$\pm$0.11&0.73$\pm$0.04&0.86$\pm$0.07&2.93$\pm$0.05&2.95$\pm$0.08  &3.02$\pm$0.05 &3.34$\pm$0.20 &0.81$\pm$0.04&0.88$\pm$0.19\\
ENb&1.86$\pm$0.12&1.35$\pm$0.10&1.00$\pm$0.05&0.87$\pm$0.05&3.47$\pm$0.05&3.10$\pm$0.06  &4.06$\pm$0.10 &3.39$\pm$0.12 &1.21$\pm$0.08&0.56$\pm$0.10\\
ENc&0.40$\pm$0.05&0.35$\pm$0.07&0.20$\pm$0.04&0.14$\pm$0.03&0.91$\pm$0.04&0.69$\pm$0.03  &0.99$\pm$0.05 &0.76$\pm$0.06 &0.22$\pm$0.04&0.16$\pm$0.05\\
W $^3$&             &             &             &1.28$\pm$0.28&              &              &4.93$\pm$0.52&              &1.08$\pm$0.41&             \\
W1c&             &             &             &1.10$\pm$0.24&              &              &2.78$\pm$0.32 &3.02$\pm$0.49 &1.13$\pm$0.26&1.32$\pm$0.41\\
W$_{ext}$ &             &             &             &2.12$\pm$0.66&7.35$\pm$0.64&6.00$\pm$0.74  &5.77$\pm$0.78 &              &3.01$\pm$0.68&             \\
C1 &1.41$\pm$0.12&1.14$\pm$0.13&0.54$\pm$0.09& 0.31$\pm$0.08 &1.55$\pm$0.10&1.52$\pm$0.09  &1.74$\pm$0.12 &1.39$\pm$0.24 &0.38$\pm$0.10&   $<$0.876 \\
C2 &0.54$\pm$0.10&0.47$\pm$0.11&0.31$\pm$0.07&   &1.59$\pm$0.09&1.27$\pm$0.06  &1.07$\pm$0.12$^2$ &0.88$\pm$0.13&          &   $<$0.465          \\
C3 &0.55$\pm$0.17&             &0.26$\pm$0.07&0.32$\pm$0.06&0.77$\pm$0.08&0.64$\pm$0.07  &1.49$\pm$0.19 &1.19$\pm$0.18 &             &0.54$\pm$0.15\\
L1 &             &0.49$\pm$0.11&0.30$\pm$0.07&0.28$\pm$0.05&3.17$\pm$0.08&2.95$\pm$0.06  &1.28$\pm$0.12 &1.29$\pm$0.12 &             & $<$0.429            \\
L2 &             &             &             &             &0.64$\pm$0.06&0.40$\pm$0.04  &              &              &             &             \\
L3 &             &             &             &             &0.70$\pm$0.09&0.77$\pm$0.05  &0.50$\pm$0.12 &0.49$\pm$0.12 &             &             \\
L4 &             &             &             &             & (0.74$\pm$0.06)$^4$&(0.31$\pm$0.05)$^4$ &0.76$\pm$0.09 &0.59$\pm$0.13 &0.29$\pm$0.08&             \\
L5 &             &             &0.71$\pm$0.10&0.41$\pm$0.05&3.16$\pm$0.12&2.56$\pm$0.06  &1.89$\pm$0.23$^2$ &1.55$\pm$0.11 &         &   $<$0.565           \\
L6 &             &             &             &             &0.61$\pm$0.08&0.59$\pm$0.06  &0.51$\pm$0.08 &0.52$\pm$0.10 &             &             \\

\hline 
\hline

\end{tabular}

\tablefoot{$^1$From the integrated spectra in the apertures defined in Fig. \ref{fig:Apertures}, in units of 10$^{-18}$ erg/s/cm$^{-2}$. For E and W galaxies we also include values in apertures of 1 arcsec in diameter as in \cite{Alvarez-MarquezSPT}. No aperture corrections, nor 10 percent correction for R100 fluxes are applied. $^2$ The fit for the H$\alpha$ -[NII] complex gives no flux for the [NII]lines. $^3$ For W, R100 H$\beta$ flux is contaminated by the lens galaxy. For the H$\alpha$ flux the R100 spectrum is preferred because the large velocity range covered by the aperture induces a significant increase of noise in the high resolution spectrum. $^4$ Difference between R100 and R2700 fluxes are larger than $\times$ 2, and not considered. The upper limits correspond to 3-$\sigma$ the noise level.}
\end{tiny}
\end{table*}


\section{Magnification map}
\label{appendix:lens_model}

The present data allow us to determine accurately the redshift, light profile, and stellar mass of the lens galaxy, and therefore they offer the opportunity to create a lens model and obtain the magnification factors for each source in the FoV. To this aim we use the public lens modeling software PyAutoLens (\citealp{Nightingale2015},  \citealp{Nightingale2018, Nightingale2021}), and follow the methodology described in \citet{Jones23}.
 
\begin{figure*}[h]
    \centering
    \includegraphics[width=0.8\textwidth]{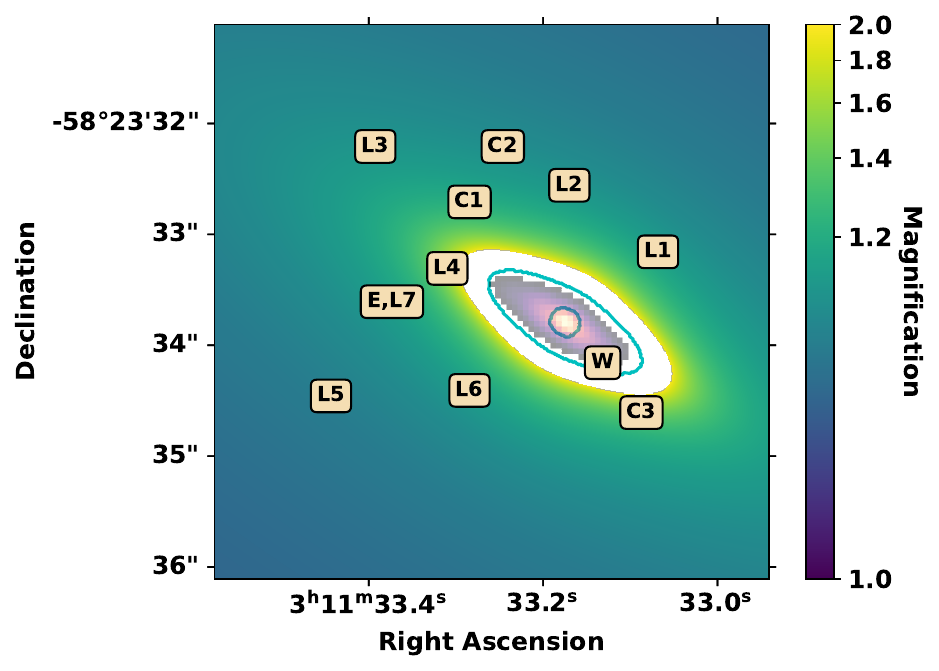}
    \caption{Magnification map defined as the inverse of the determinant of the Jacobian of our best-fit lens model. The critical lines are displayed in cyan, and positions of galaxies are marked with text. In order to display the small magnifications of the distant components, we only display magnifications below $\mu<2$. For reference, we display the panchromatic image of the lensing galaxy. 
    } 
    \label{fig:lens_magnification}
\end{figure*}

First, we generate a panchromatic image collapsing the R100 datacube over the spectral range 1--5\,$\mu$m (i.e. rest-frame $\sim$ 0.49 -- 2.45\,$\mu$m at the lens redshift). The light profile for the lens galaxy in this image was fitted with a S\'{e}rsic profile, and we model the PSF as a circular Gaussian with FWHM$= 0.1''$. The fitting process results in the best-fit centroid position, intrinsic axis ratio ($q\equiv a/b$), position angle ($\phi$), effective radius ($\rm r_{eff}$), and S\'{e}rsic index (n) of the light profile. We assume that the mass profile can be represented by an isothermal ellipsoid with the same geometrical properties as the light profile (i.e. centroid, q, $\phi$), but an Einstein radius based on the stellar mass of the galaxy obtained from the R100 spectra as in Sect. \ref{subsec:Stellar masses} (i.e. log(\Mstar/M$_\odot$ = 9.39) and the expected stellar mass surface fraction following \cite{Jimenez-Vicente2015}. The resulting lens parameters are listed in Table \ref{lens_properties}. This model was used to create a magnification map, from which we obtained the factors included in Table \ref{mag_factors} (see also Fig. \ref{fig:lens_magnification}). Since source W is multiply imaged, it was not included in this simple analysis, and we adopt the magnification factor of \cite{Marrone2018} instead. 

\begin{table*}
\centering
\begin{tabular}{ccccc}
$q$   & $\phi$ [$^{\circ}$]       & $\rm r_{eff}$ [$''$] & $n$ & $\rm r_{Ein}$ [$''$]\\ \hline
$2.92\pm0.01$ & $59.32\pm0.08$ & $0.266\pm0.001$ & $1.01\pm0.01$ & $0.17$
\end{tabular}
\caption{Best-fit lens model parameters, including axis ratio ($q$), position angle $\phi$, effective radius and S\'{e}rsic radius of the light profile, and Einstein radius.}
\label{lens_properties}
\end{table*}

\begin{table}
\begin{tabular}{c|c}
Component & Magnification \\ \hline
E,L7 & 1.15 \\
C1 & 1.21 \\
C2 & 1.10 \\
C3 & 1.40 \\
L1 & 1.14 \\
L2 & 1.12 \\
L3 & 1.11 \\
L4 & 1.62 \\
L5 & 1.06 \\
L6 & 1.14 \\
W & (2.2) \\
\end{tabular}
\caption{Approximate magnification factors for each galaxy in the FoV, as derived from the determinant of the Jacobian of our best-fit lens model. Since source W is multiply imaged, it is not included in this simple analysis and we state the magnification factor of \citet{Marrone2018} instead.}
\label{mag_factors}
\end{table}

\end{appendix}

\end{document}